\documentclass{JHEP}

\usepackage{amssymb}

\usepackage{amsfonts}

\usepackage{amsbsy}

\usepackage{epsfig}

\newcommand{\im}{\mathop{\rm Im}\nolimits}
\newcommand{\re}{\mathop{\rm Re}\nolimits}
\newcommand{\NP}[1]{Nucl.\ Phys.\ {\bf #1}}

\renewcommand{\cal}{\mathcal}

\def\IC{\mathbb{C}}
\def\IZ{\mathbb{Z}}
\def\IR{\mathbb{R}}

\def\sd{\boldsymbol{d}}

\def\sx{{\bf x}}

\def\su{{\bf u}}
\def\sv{{\bf v}}

\def\CK{{\cal K}}

\def\N{{\cal N}}
\def\bb{\bar{b}}
\def\bz{\bar{z}}
\def\bZ{\bar{Z}}

\def\bpartial{\bar{\partial}}

\def\s*{\boldsymbol{*}}

\def\mod{\mbox{ mod }}

\newcommand{\be}{\begin{equation}}
\newcommand{\ee}{\end{equation}}
\newcommand{\bea}{\begin{eqnarray}}
\newcommand{\eea}{\end{eqnarray}}

\def\Om{\Omega_0}
\def\Omn{\Omega}
\def\CM{\cal M}

\title{(Dis)assembling Special Lagrangians}

\author{Frederik Denef \\ Department of Mathematics,
Columbia University \\ New York, NY 10027, USA\\
\email{denef@math.columbia.edu}}

\preprint{{\tt hep-th/0107152}}

\abstract{We explain microscopically why split attractor flows,
known to underlie certain stationary BPS solutions of four
dimensional $\N=2$ supergravity, are the relevant data to describe
wrapped D-branes in Calabi-Yau compactifcations of type II string
theory. We work entirely in the context of the classical geometry
of A-branes, i.e.\ special Lagrangian submanifolds, avoiding both
the use of homological algebra and explicit constructions of
special Lagrangians. Our results provide a way to disassemble and
assemble arbitrary special Lagrangians to and from more simple
building blocks, giving a concrete way to determine for example
marginal stability walls and deformation moduli spaces.}

\begin{document}

\section{Introduction}\label{sec:intro}
\setcounter{equation}{0}

There are many good reasons to study D-branes on Calabi-Yau
compactifications: they provide a nontrivial testing ground for
virtually all ideas in D-brane physics, combine a plethora of
beautiful results in mathematics and physics, and promise new key
insights in the dynamics of $\N=1$ gauge theories (for a nice
short review, see \cite{Dstrings}).

One distinguishes two types of D-branes in type II Calabi-Yau
compactifications, B-branes, wrapping holomorphic cycles and
carrying holomorphic bundles, and A-branes, wrapping special
Lagrangian submanifolds and carrying flat connections. A- and
B-branes are allegedly interchanged by mirror symmetry.

Though A-branes have a much more obvious geometrical framework to
be analyzed in than B-branes, the main focus of research
\cite{BDLR,DFR,Dcat,Dstrings,prevwork} has been on B-branes, in
part because it is virtually impossible to construct generic
special Lagrangians explicitly, and in part because some quantum
corrections are better in control for B-branes. However, studying
B-branes properly requires the introduction of some arcane --- and
for most of us even scary --- mathematics. The state of the art
can be found in \cite{Dcat}, briefly summarized in
\cite{Dstrings}.

In this paper on the other hand, we will exclusively consider
A-branes, without quantum or stringy corrections, that is, the
geometry of special Lagrangian submanifolds. Other work dealing
directly with special Lagrangians includes
\cite{joyce,hitchin,SYZ,KM,Th,ThY,LazSLG}. As it turns out, we
will only need some pretty basic mathematics, enhanced with
relatively well-known features of special geometry. This hopefully
will take away some of the barriers complicating access to this
field of research.

The main direct motivation for this work was the observation in
\cite{branessugra,DGR} that split attractor flows on the vector
multiplet moduli space provide a remarkably powerful way to
analyze existence and stability of BPS states in type II
compactifications. Such flows appear naturally as the basic data
describing BPS solutions of the corresponding four dimensional
$\N=2$ supergravity theory, but turn out to provide accurate
predictions even beyond the regime where supergravity can be
expected to be valid. Also the exact same structures, in a
completely different guise, had turned up before in the context of
$\N=2$ field theories (decoupled from gravity), as the appropriate
data to desribe BPS states \cite{BPS37}. The obvious question all
this prompted was whether the appearance of split flows could also
be understood from an entirely microscopic D-brane perspective. In
view of the many known examples of identical basic data describing
BPS objects in widely different regimes, a positive answer was
certainly not unlikely.

We will provide that missing link in what follows. The main idea
is that split attractor flows can be associated to certain
deformations of special Lagrangians (SLGs), providing a way to
assemble and disassemble these from and to more simple building
blocks. This in turn gives substantial insight in the structure of
arbitrary SLGs, yielding for instance their domain of stability
and in favorable circumstances a parametrization of their
(uncorrected) moduli spaces. Our scheme does not require explicit
construction of generic SLGs, which, in view of the effective
impossibility of such constructions, is obviously rather good
news.

The outline of the paper is as follows. In section \ref{sec2}, we
recall some well-known and less well-known features of type IIB
Calabi-Yau compactifications, including a few useful facts about
special Lagrangians and their deformations. We make the latter
quite explicit, thus preparing ourselves for section
\ref{sec:micrattr}. In section \ref{sec:properties}, we briefly
review (split) attractor flows. Though originally introduced in
the context of supergravity, we will reformulate things in a way
which no longer refers directly to spacetime structures. Section
\ref{sec:micrattr} forms the core of this paper. It explains how
changing the complex structure along an attractor flow induces a
certain deformation of any associated special Lagrangian $L$,
determined by requiring preservation of the SLG condition. Such
deformations can cause $L$ to split. We investigate in detail the
different possible degenerations of this kind, with particular
emphasis on determining the ``direction'' of decay, and link them
in a natural way to the different kinds of splits observed in the
attractor flow picture. In section \ref{sec:decomp}, we use all
this to construct a procedure for splitting up arbitrary special
Lagrangians into simpler ones, by letting them flow, according to
the rules of section \ref{sec:micrattr}, along attractor flow
trees. Reversing this procedure gives in turn a way to assemble
special Lagrangians of arbitrary complexity out of simple building
blocks. Explicit construction of these special Lagrangians is not
required; the flow trees themselves already encode a great deal of
very useful information (e.g.\ stability). In this manner we
obtain an efficient classification scheme for special Lagrangians.
We end the section with a comparison with the $\Pi$-stability
conjecture. Finally, in section \ref{sec:examples}, we illustrate
some of the general results with a couple of simple examples. The
reader might find it useful to have a look at these already while
reading the previous sections.

Some of the results derived in this paper have been obtained
before by mathematicians. The simplest SLG splittings in section
\ref{sec:micrattr} were first studied by Joyce \cite{joyce}.
However, the methods we use, based on infinitesimal deformations
of arbitrary compact SLGs (and a useful physical analogy with
steady heat flow) rather than local models, are more concrete and
perhaps more general, at least from the point of view of a
physicist. Secondly, the general concept of studying Lagrangians
through Hamiltonian deformations, the importance of various
connect sums in this context, and the notion of a Lagrangian
decomposition appeared already in the recent work of Thomas
\cite{Th} and Thomas and Yau \cite{ThY}. The immediate scope of
their work is quite different from ours though. Roughly, in
physical terms, they consider (among other things) D-branes
wrapped around \emph{arbitrary} Lagrangians, and investigate
whether these will ``decay'', at \emph{fixed} complex structure,
to one or more branes wrapped around \emph{special} Lagrangians,
where the decay process is taken to be given by the mean curvature
Hamiltonian flow acting on the Lagrangian till it splits, and then
further on its decay products. In the present paper on the other
hand, we consider only \emph{special} Lagrangians, and propose a
way to (dis)assemble those by letting the complex structure
\emph{vary} along an attractor flow tree. Physically, the
disassembling process can be thought of as adiabatically moving
the special Lagrangian into its corresponding large $N$
supergravity solution. So the two setups are inherently different,
though (at least in simple cases) some obvious relations exist.

\section{Some (less) well-known features of type IIB Calabi-Yau
compactifications} \label{sec2}

For concreteness, we will work in the framework of type IIB string
theory compactified on a Calabi-Yau 3-fold $X$. This theory has
$\N=2$ supersymmetry in four dimensions, with $n_v = h^{1,2}(X)$
massless abelian vector multiplets and $n_h = h^{1,1}(X)+ 1$
massless hypermultiplets. The vector multiplet scalars are given
by the complex structure moduli of $X$, and the lattice of
electric and magnetic charges is identified with $H^3(X,\IZ)$, the
lattice of integral harmonic $3$-forms on $X$: after a choice of
symplectic basis ${\alpha^I,\beta_I}$ of $H^3(X,\IZ)$, a D3-brane
wrapped around a cycle Poincar\'e dual to $\Gamma \in H^3(X,\IZ)$
has electric and magnetic charges equal to its components with
respect to this basis.

\subsection{Special Geometry of the complex structure moduli space}

The geometry of the complex structure moduli space ${\cal M}_c$,
parametrized by $n_v$ coordinates $z^a$, is special
K\"ahler~\cite{SG}. The (positive definite) metric
\begin{equation} \label{SKmetric}
 g_{a\bb} = \partial_a \bpartial_{\bb} \CK
\end{equation}
is derived from the K\"ahler potential
\begin{equation} \label{kahlerpotential}
\CK = - \ln \left( i \int_X \Om \wedge \overline{\Om} \right),
\end{equation}
where $\Om$ is the holomorphic $3$-form on $X$, depending
holomorphically on the complex structure moduli. It is convenient
to introduce also the {\em normalized} 3-form\footnote{In
\cite{DGR}, the holomorphic 3-form was denoted as $\Omega$, and
the normalized one as $\tilde{\Omega}$.}
\begin{equation} \label{Omdef}
\Omn  \equiv e^{\CK/2} \, \Om\,.
\end{equation}
Then the ``central charge'' of $\Gamma \in H^3(X,\IZ)$ is given by
\begin{equation} \label{Zdef}
Z(\Gamma) \equiv \int_X \Gamma \wedge \Omn \equiv \int_\Gamma
\Omn\,,
\end{equation}
where we denoted, by slight abuse of notation, the homology class
Poincar\'e dual to $\Gamma$ by the same symbol $\Gamma$. Note that
$Z(\Gamma)$ has a nonholomorphic dependence on the moduli through
the K\"{a}hler potential. It can be shown \cite{BBS,D0} that for
any three (real) dimensional submanifold $L$ of $X$,
$\mathrm{Vol}(L) \geq k |Z(L)|$, where $k$ is a constant
independent of the complex structure moduli. More precisely,
$k=\sqrt{8 \mathrm{Vol}(X)}$. Equality is obtained if $L$ is
special Lagrangian (see below). The mass of the corresponding
wrapped BPS 3-brane is $M=|Z(L)|/\sqrt{G_N}$, with $G_N$ the four
dimensional Newton constant.

Central in what follows will be the (antisymmetric, topological,
moduli independent) \emph{intersection product}, defined as:
\begin{equation} \label{intproddef}
\langle \Gamma_1,\Gamma_2 \rangle = \int_X \Gamma_1 \wedge
\Gamma_2 = \int_{\Gamma_1} \Gamma_2 = \#(\Gamma_1 \cap
\Gamma_2)\,,
\end{equation}
where the intersection points are counted with signs. With this
notation, we have for a symplectic basis $\{ \alpha^I,\beta_I \}$
by definition $\langle \alpha^I,\beta_J \rangle = \delta^I_J$, so
for $\Gamma_i = q_i^I \beta_I - p_{i,I} \alpha^I$, we have
$\langle \Gamma_1,\Gamma_2 \rangle = q_1^I p_{2,I} - p_{1,I}
q_2^I$. This is nothing but the Dirac-Schwinger-Zwanziger
symplectic inner product on the electric/magnetic charges.
Integrality of this product is equivalent with Dirac charge
quantization.

Every harmonic $3$-form $\Gamma$ on $X$ can be decomposed
according to $H^3(X,\IC) = H^{3,0}(X) \oplus H^{2,1}(X) \oplus
H^{1,2}(X) \oplus H^{0,3}(X)$ as (for real $\Gamma$):
\begin{equation} \label{Gamdecomp}
 \Gamma = i \bZ(\Gamma) \, \Omn \, - \, i g^{a\bb} \bar{D}_{\bb} \bar{Z}(\Gamma) \,
 D_a \Omn \, + \mbox{ c.c.} \, ,
\end{equation}
where $D_a \equiv (\partial_a + \frac{1}{2} \partial_a\CK)$.
This decomposition is orthogonal with respect to the
intersection product (\ref{intproddef}).

\subsection{Special Lagrangian submanifolds and their
deformations}\label{sec:SLGdeform}

At large volume and zero string coupling, the condition for a
single wrapped D3-brane to be supersymmetric is that it is
embedded (or, to be precise, immersed) in the Calabi-Yau manifold
$X$ as a \emph{special Lagrangian} (SLG) submanifold, and that the
$U(1)$ gauge field on its worldvolume is flat \cite{BBS}. A three
real dimensional submanifold $L$ of $X$ is called special
Lagrangian with phase $\alpha$ if
\begin{eqnarray}
 \omega \,|_L &=& 0 \\
 \im(e^{-i \alpha} \Omega) \,|_L &=& 0
\end{eqnarray}
and its orientation is such that $\int_{L} e^{-i \alpha} \Omega >
0$. Here $|_L$ denotes the pull-back to $L$, and $\omega$ is the
K\"ahler form on $X$. Then $\int_{L} e^{-i \alpha} \Omega \, |_L$
is up to a constant factor equal to the volume form on $L$, and as
stated earlier, the volume of $L$ saturates the BPS bound:
\begin{equation} \label{VolSLG}
 \mathrm{Vol}(L)=k |Z(L)| \, ,
\end{equation}
with $k=\sqrt{8 \mathrm{Vol}(X)}$.

It can be shown \cite{hitchin,SYZ} that the moduli space of
deformations of $L$ has real dimension $b^1(L) = \dim H^1(L,\IR)$,
its tangent space at $L$ being isomorphic to $H^1(L,\IR)$, the
space of real harmonic 1-forms on $L$. On the other hand, there
are also $b^1(L)$ moduli corresponding to Wilson lines of the flat
$U(1)$ gauge field. The deformations pair up with these moduli to
form the $b^1(L)$ complex dimensional D-brane moduli space.

The correspondence between SLG deformations and harmonic 1-forms
can be made explicit as follows.\footnote{The calculation given
here follows the reasoning outlined in \cite{hitchin,SYZ}. We are
somewhat more explicit here, preparing for section
\ref{sec:micrattr}.} Let $L$ be a smooth SLG, $I$ some open
interval containing $0$ and $f_t:L \to X$, $t \in I$, a one
parameter family of arbitrary smooth deformations of $L$, with
$f_0(L)=L$. Define the map $F:I \times L \to X$ by $F(t,\sx)
\equiv f_t(\sx)$. Then we can write
\begin{eqnarray}
 F^* \omega &=& \theta^{(1)} \wedge dt + \sigma^{(2)} \label{slgflow1} \\
 F^* \im(e^{-i \alpha} \Omn) &=& \eta^{(2)} \wedge dt + \chi^{(3)}
 \, , \label{slgflow2}
\end{eqnarray}
where the various Greek letters denote various $t$-dependent
differential forms on $L$. Note that the deformations preserve the
SLG condition, that is, ${f_t}^* \omega = 0$ and ${f_t}^*
\im(e^{-i \alpha} \Omn)=0$ for all $t$, if and only if
$\sigma^{(2)} = 0$ and $\chi^{(3)}=0$.

We would like to find an equivalent condition for SLG
preservation, but now on the forms $\theta^{(1)}$ and
$\eta^{(2)}$. This goes as follows. Since $d \, F^* \omega = F^*
d\omega = 0$ and $d \, F^* \im(e^{-i \alpha} \Omn) = F^* \im(e^{-i
\alpha} d \Omn) = 0$, equations (\ref{slgflow1})-(\ref{slgflow2})
imply:
\begin{eqnarray}
 0 &=& \sd \theta^{(1)} \wedge dt + \partial_t \sigma^{(2)} \wedge dt + \sd \sigma^{(2)} \label{dslgflow1} \\
 0 &=& \sd \eta^{(2)} \wedge dt - \partial_t \chi^{(3)} \wedge dt +
 \sd \chi^{(3)} \, , \label{dslgflow2}
\end{eqnarray}
where $\sd$ denotes the exterior derivative on $L$. Therefore, the
SLG condition is preserved by the deformations if and only if
\begin{equation} \label{etclosed}
 \sd \theta^{(1)} = 0 = \sd \eta^{(2)} \, .
\end{equation}

Writing out components, one sees that the definition
(\ref{slgflow1})-(\ref{slgflow2}) for these forms is equivalent
(for arbitrary deformations) with the following explicit
expressions, in obvious notation:
\begin{eqnarray}
 \theta^{(1)} &=& 2 \im [g_{m\bar{n}} \, \partial_t F^m \, \partial_i {\bar
 F}^{\bar{n}}] \, dx^i \\
 \eta^{(2)} &=& 3 \im[e^{-i \alpha} \Omn_{mnr} \, \partial_t F^m
 \, \partial_i F^n \, \partial_j F^r] \, dx^i \wedge dx^j \, .
\end{eqnarray}
($i,j,\ldots$ are real coordinate indices on $L$ and
$m,n,r,\ldots$ holomorphic coordinate indices on $X$, with
$g_{m\bar{n}}$ the Ricci-flat metric.) Putting $t=0$ and
decomposing the deformation vector field $\partial_t F \,|_{t=0}$
on $L$ in a tangential\footnote{The longitudinal part $w$ can
always be put to zero; it is simply the gauge degree of freedom
for diffeomorphisms of $L$.} ($w$) and a normal ($v$) part as
\begin{equation}
 \partial_t F^m \,|_{t=0} \equiv (w^j + i \, v^j)
\partial_j F^m \,|_{t=0} \, ,
\end{equation}
and using the SLG properties of $L$, this can be rewritten as
\begin{eqnarray}
 \theta^{(1)} &=& 2 h_{ij} v^i \, dx^j \label{tcomp} \\
 \eta^{(2)} &=& \frac{1}{2 k} \sqrt{h} \, \epsilon_{ijk} \, v^i \, dx^j
 \wedge dx^k \, , \label{ecomp}
\end{eqnarray}
where $h_{ij}$ is the induced metric on $L$ and, as before,
$k=\sqrt{8 \mathrm{Vol}(X)}$ (the constant $k$ appears here
because we used equation (\ref{VolSLG})). Thus we get a one-to-one
correspondence between (arbitrary) infinitesimal deformations and
1- or 2-forms on $L$. It is furthermore straightforward now to
check that
\begin{equation}
 * \theta^{(1)} = 2 k \, \eta^{(2)} \, .
\end{equation}
Combining this with the condition (\ref{etclosed}) for \emph{SLG}
deformations, we see that $\theta^{(1)}$ corresponds to an
infintesimal SLG deformation if and only if it is harmonic,
yielding the isomorphism between the tangent space to the moduli
space of SLG manifolds at $L$ and $H^1(L,\IR)$, as announced.

\section{Attractor flows and their (not so) basic properties}\label{sec:properties}
\setcounter{equation}{0}

\subsection{Definition}

Attractor flows have their origin in the description of black hole
solutions of four dimensional $\N=2$ supergravity theories
\cite{FKS,attrs,M,branessugra}. Here we will simply define them by
a certain flow equation in moduli space.\footnote{In the 4d
supergravity context, these flow equation usually involve also the
space-dependence of the metric. However, this is easily eliminated
\cite{M}, leaving pure flow equations in moduli space.} The data
specifying a flow are a cohomology class (or charge) $\Gamma \in
H^3(X,\IZ)$ and an initial point $z_0$ in the complex structure
moduli space $\CM_c$. The corresponding attractor flow is an
oriented trajectory in $\CM_c$ given by the solution of
\begin{equation}
 \mu \, \frac{d z^a}{d \mu} = g^{a\bb} \, \bpartial_{\bb} \ln |Z|^2 \,, \label{at}
\end{equation}
where $\mu$ is always positive and runs \emph{down} starting from
$1$, $z_{\mu=1}=z_0$, $Z=Z(\Gamma)$ as in (\ref{Zdef}) and
$g_{a\bb}$ is the special K\"ahler metric (\ref{SKmetric}) on
$\CM_c$.\footnote{In the 4d black hole context, the physical
meaning of $\mu$ is the metric redshift factor: $ds^2=-\mu^2 dt^2
+ \mu^{-2} d\sx^2$. The spatial dependence of $\mu$ is then given
by $\partial_\tau \mu = -\mu^2 |Z|$, with $\tau=1/|\sx|$.
\label{fnBH}} Thus attractor flows are simply gradient lines of
the potential $\ln |Z|^2$. In particular, $|Z|$ decreases with
decreasing $\mu$, hence the flow will tend to minimal $|Z|$.
Generically, this local minimum of $|Z|$ is isolated, so the
endpoint of the flow, the so-called attractor point, is invariant
under small variations of the starting point $z_0$. It can be
shown \cite{M} that all critical points of $\ln |Z|^2$ are in fact
local minima.

If the relevant local minimum of $|Z|$ is nonzero, the flow is
smooth and $\mu$ runs all the way down to 0. If on the other hand
the minimum is zero, this zero will generically be hit before
$\mu=0$, and the flow stops there. We will see later on that there
is a further natural distinction between regular zeros of $Z$ and
zeros at singularities in $\CM_c$ where the cycle $\Gamma$
collapses.\footnote{Attractor points with nonzero minimal $|Z|$
correspond to spherically symmetric black holes with finite
horizon area, regular zeros of $Z$ correspond to charges that
don't admit a spherically symmetric BPS solution, and zeros at
singularities can either correspond to singular black holes with
zero horizon area, or horizonless enhan\c{c}on-like ``empty
holes'' \cite{M,branessugra,enhancon,argyres}.}

It will be crucial in what follows to know how $\im(e^{-i \alpha}
\Omn)$, with $\alpha \equiv \arg Z$, varies along the attractor
flow. A short calculation shows that for an infinitesimal change
in complex structure $\delta z^a$ along the attractor flow, we
have $\delta \alpha = - \im(\delta z^a \partial_a \CK)$, so
\begin{equation} \label{varom1}
 \delta \, \im (e^{-i \alpha} \Omn ) = \im (e^{-i \alpha} \, \delta z^a D_a
 \Omn)\, .
\end{equation}
Combining this with (\ref{at}) and the identity $\partial_{\bb}
\ln |Z|^2 = \bar{D}_{\bb} \bar{Z} / \bar{Z}$ gives
\begin{equation}
 \mu \, \frac{d}{d\mu} \, \im (e^{-i \alpha} \Omn ) =
 \frac{1}{|Z|}\, \im(g^{a\bb}\, \bar{D}_{\bb} \bar{Z} \, D_a \Omn)
 \, ,
\end{equation}
which, using (\ref{Gamdecomp}), can be elegantly rewritten as
\begin{equation} \label{integrated}
\mu \, \frac{d}{d\mu} \, \im (e^{-i \alpha} \Omn ) - \im (e^{-i
\alpha} \Omn ) = \frac{\Gamma}{2 |Z|} \, .
\end{equation}
Since $\Gamma$ is constant in cohomology, this equation can be
integrated if we consider it as an equation in $H^3(X,\IC)$:
\begin{equation} \label{integrated2}
  2 \mu^{-1} \im(e^{-i \alpha} \Omn) = - \Gamma \, \tau
  + 2 \im(e^{-i \alpha} \Omn)_0.
\end{equation}
where $\tau(\mu)$ is defined\footnote{In the supergravity picture,
$\tau$ appears naturally, as in footnote \ref{fnBH}.} by $|Z| d
\tau = - d\mu / \mu^2$, $\tau_{\mu=1}=0$, and where, as in the
remainder of the paper, the index ``0'' refers to the initial
point $z_0$. Note that when $\mu$ runs down from 1 to 0, $\tau$
runs up from 0 to infinity.

Finally, for completeness, we recall that attractor flows can also
be considered as geodesic strings \cite{DGR} with respect to the
action
\begin{equation} \label{stringaction}
 S=|Z_*| + \int 2 \sqrt{g^{a\bb}
\partial_a |Z| {\bpartial}_{\bb} |Z|} \, \, ds \, ,
\end{equation}
where the starting point of the string is kept fixed at $z_0$,
$Z_*$ is $Z(\Gamma)$ evaluated at the free endpoint of the string,
and $ds$ is the line element on $\CM_c$ : $ds^2=g_{a\bb} dz^a
d\bz^{\bb}$. Requiring $\delta S=0$ for variations of the free
endpoint fixes the latter to be located at the attractor point of
$\Gamma$. The action $S$ reaches its minimal value, equal to
$|Z|_0$, when evaluated along an attractor flow.

For practical techniques to solve attractor flow equations in this
formalism, we refer to \cite{DGR}.

\subsection{Marginal Stability} \label{sec:ms}

A central concept in the discussion of stability of BPS states
(and therefore of SLGs) is \emph{marginal stability} (MS).
Generically, a BPS state of charge $\Gamma$ is stable against the
decay $\Gamma \to \Gamma_1 + \Gamma_2$ by conservation of energy
and the triangle inequality: $M = |Z| = |Z_1 + Z_2| \leq |Z_1| +
|Z_2| \leq M_1 + M_2$. However, if the decay products $\Gamma_1$
and $\Gamma_2$ are BPS, and $\arg Z_1 = \arg Z_2 (= \arg Z)$, the
state is only \emph{marginally} stable against this decay, i.e.\
the decay is no longer forbidden by conservation of energy. A
typical locus where $\arg Z_1 = \arg Z_2$ is of codimension one,
and is called a wall (or hypersurface) of $(\Gamma_1,\Gamma_2)$
marginal stability.\footnote{Also often called a \emph{line} of
marginal stability, in analogy with the most famous example
\cite{SW}, where this locus is indeed a line (and moreover unique,
another rather atypical feature).} Upon crossing such a wall, it
is possible (though not necessary) that certain one particle BPS
states with charge $\Gamma$ are forced to decay into certain two
particle states that are no longer BPS (consisting of the charges
$\Gamma_1$ and $\Gamma_2$). The analog of this for SLGs is the
Joyce transition, where a single SLG splits into two SLGs with
different phases \cite{joyce,KM} (see section \ref{sec:micrattr}).

\FIGURE[t]{\centerline{\epsfig{file=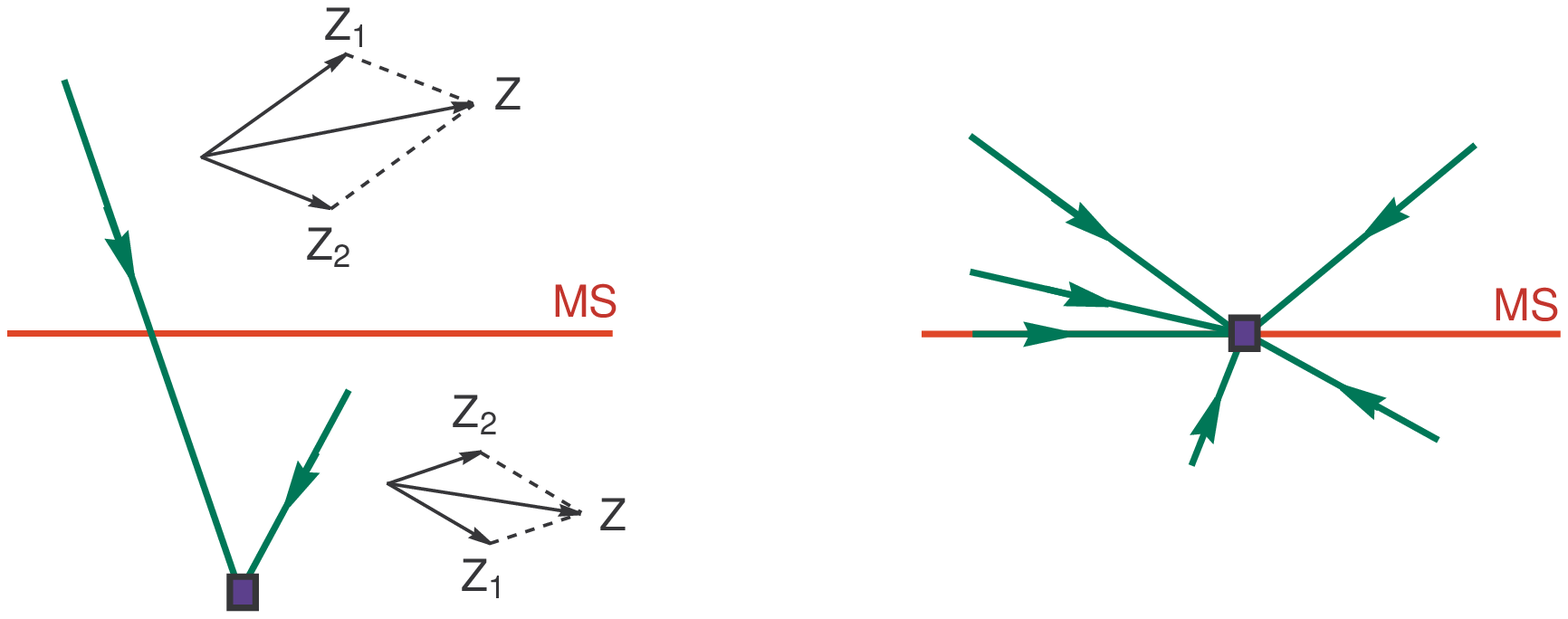,height=6cm}}
\caption{\emph{Left:} behavior of attractor flows near a
$(\Gamma_1,\Gamma_2)$ MS line with $\langle \Gamma_1,\Gamma_2
\rangle > 0$. Typical central charge phases corresponding to
intersection or no intersection are indicated. \emph{Right:}
behavior of attractor flows near a $(\Gamma_1,\Gamma_2)$ MS line
with $\langle \Gamma_1,\Gamma_2 \rangle = 0$.}\label{msint}}

It will be important for us to know whether or not an attractor
flow with charge $\Gamma = \Gamma_1 + \Gamma_2$ can cross or reach
a wall of marginal stability for the decay $\Gamma \to \Gamma_1 +
\Gamma_2$. This is easily answered by taking the intersection
product of equation (\ref{integrated2}) with $\Gamma_1$. This
gives, using $\langle \Gamma_1,\Gamma \rangle = \langle \Gamma_1,
\Gamma_2 \rangle$:
\begin{equation} \label{Z1eq}
2 \im [e^{-i \alpha} Z_1 ] = - \langle \Gamma_1, \Gamma_2 \rangle
\, \mu \, \tau + 2 \mu \im [e^{-i \alpha} Z_1 ]_0 \, .
\end{equation}
By definition, at marginal stability, the left hand side is zero.
Now there are two cases to distinguish, as illustrated in fig.\
\ref{msint}. The first (right hand side in the figure) is when
$\Gamma_1$ and $\Gamma_2$ have zero intersection
product.\footnote{In the four dimensional theory, this corresponds
to mutually local charges, i.e. charges that (possibly after an
electromagnetic duality rotation) can simultaneously be considered
to be electric, without magnetic components.} Putting the right
hand side of (\ref{Z1eq}) equal to zero then gives $\mu \im [e^{-i
\alpha} Z_1 ]_0 = 0$, that is, \emph{either} we are already at
$(\Gamma_1,\Gamma_2)$ marginal stability from the beginning (at
$z_0$), and then the flow stays inside the MS wall, \emph{or} we
start away from the wall, and then we can only reach it at
$\mu=0$, the attractor point. So, in particular, the flow can
never intersect the wall transversally. Note that we do not
\emph{necessarily} reach $(\Gamma_1,\Gamma_2)$ marginal stability
at the attractor point: even though, for a flow converging to a
nonzero minimal $|Z|$, we will always have $\mu=0$ at the
attractor point and therefore $\im [Z_1 \bar{Z_2}]=0$, it is still
possible (and in fact more generic) to have $\arg Z_1 = \arg Z_2
\pm \pi$ rather than equal phases.

The second case is $\langle \Gamma_1,\Gamma_2 \rangle \neq 0$.
Taking the intersection product of $\Gamma_1$ with equation
(\ref{integrated}) then implies that, at an attractor point with
$\mu=0$, we can never be at a $(\Gamma_1,\Gamma_2)$ MS wall.
However, again by intersecting (\ref{integrated2}) with
$\Gamma_1$, one sees that now the flow can intersect the wall
transversely, namely at $\tau_{ms} = 2 \im [e^{-i \alpha} Z_1 ]_0
/ \langle \Gamma_1, \Gamma_2 \rangle$, or written more
symmetrically, at
\begin{equation} \label{tms}
 \tau_{ms} = \left. \frac{2 \im (Z_1 \bar{Z}_2)}{|Z| \langle \Gamma_1,
\Gamma_2 \rangle} \, \right|_0 \, .
\end{equation}
A necessary condition for the intersection point to exist is of
course $\tau_{ms} > 0$, that is,
\begin{equation} \label{prejoycecondition}
 \langle \Gamma_1,\Gamma_2 \rangle \im(Z_1 \bar{Z}_2)_0 > 0
\end{equation}
This is also indicated in fig.\ \ref{msint}. Note that this
condition is not sufficient to have intersection with the marginal
stability wall: the flow could instead cross an ``anti-MS'' wall,
i.e.\ where $\arg Z_1 = \arg Z_2 \pm \pi $, or it could hit a zero
(also necessarily on an anti-MS wall) before reaching
$\tau=\tau_{ms}$.

\subsection{Split flows} \label{sec:splitflows}

\FIGURE[t]{\centerline{\epsfig{file=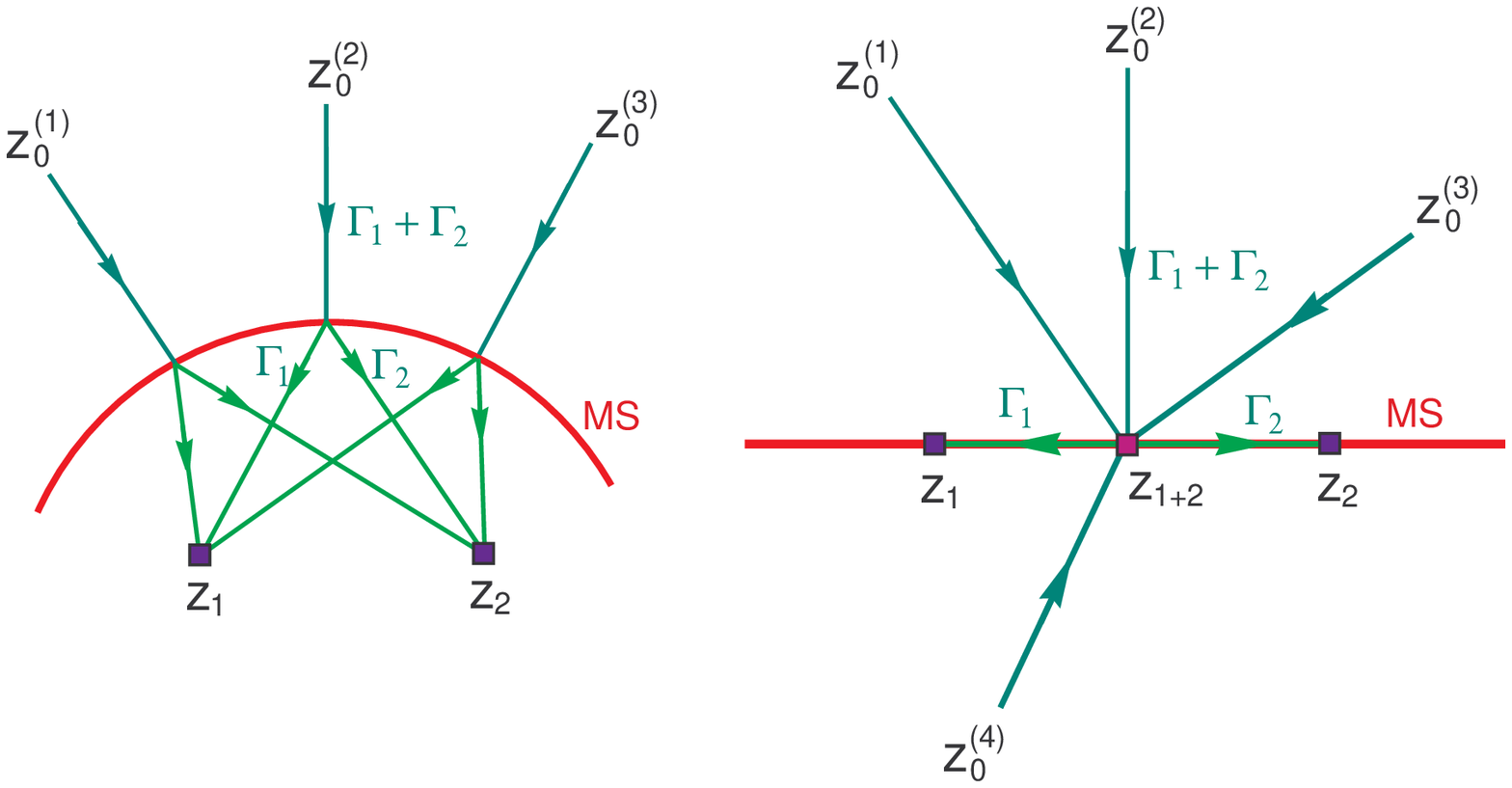,height=7cm}}
\caption{\emph{Left:} type 1 split flows. \emph{Right:} type 2
split flows. In both cases a set of initial points is indicated by
$z_0^{(i)}$, and the two final attractor points by $z_1$ and
$z_2$. For the type 2 split, the ``intermediate''
$\Gamma_1+\Gamma_2$-attractor point is denoted by
$z_{1+2}$.}\label{splits}}

Split attractor flows arise naturally in the description of
non-spherically symmetric stationary BPS supergravity solutions
\cite{branessugra}. Again we will take a pragmatic approach here
and simply state a definition, without reference to supergravity
equations of motion. A split flow is obtained by letting a flow of
charge $\Gamma = \Gamma_1 + \Gamma_2$ split, at a
$(\Gamma_1,\Gamma_2)$ MS wall, into a flow of charge $\Gamma_1$
and a flow of charge $\Gamma_2$. In view of the previous
discussion, there are two cases to consider: splits at a MS wall
with $\langle \Gamma_1,\Gamma_2 \rangle \neq 0$ (\emph{type 1})
and splits at a MS wall with $\langle \Gamma_1,\Gamma_2 \rangle =
0$ (\emph{type 2}). From the conclusions of the previous
discussion, it follows that type 1 splits can happen anywhere on
the MS line (and the position will in general be dependent on the
starting point of the flow), whereas type 2 splits necessarily
occur \emph{at} an attractor point (independent of starting
point).\footnote{In the four dimesnional supergravity, type 1
splits are related to multicentered configurations at the
macroscopic level, whereas type 2 splits are related to
multicentered configurations in the near horizon region
\cite{DGR}.} Because of that, for generic initial $z_0$, type 1
splits involve just two branches, whereas type 2 splits might
generically involve more than two. Furthermore, branches coming
out of a type 2 split necessarily stay on their MS line, while
those coming out of a type 1 split necessarily leave that line.

The two types of splits are illustrated in fig.\ \ref{splits}. The
splitting can be repeated on the separate branches, resulting in
flow trees of arbitrary complexity, as shown in fig.\
\ref{flowtree}. These split flow trees can still be considered to
be geodesic strings with respect to the action
(\ref{stringaction}), where the condition for the splits to occur
only at MS is moreover an automatic consequence of the
minimization of the action.

\FIGURE[t]{\centerline{\epsfig{file=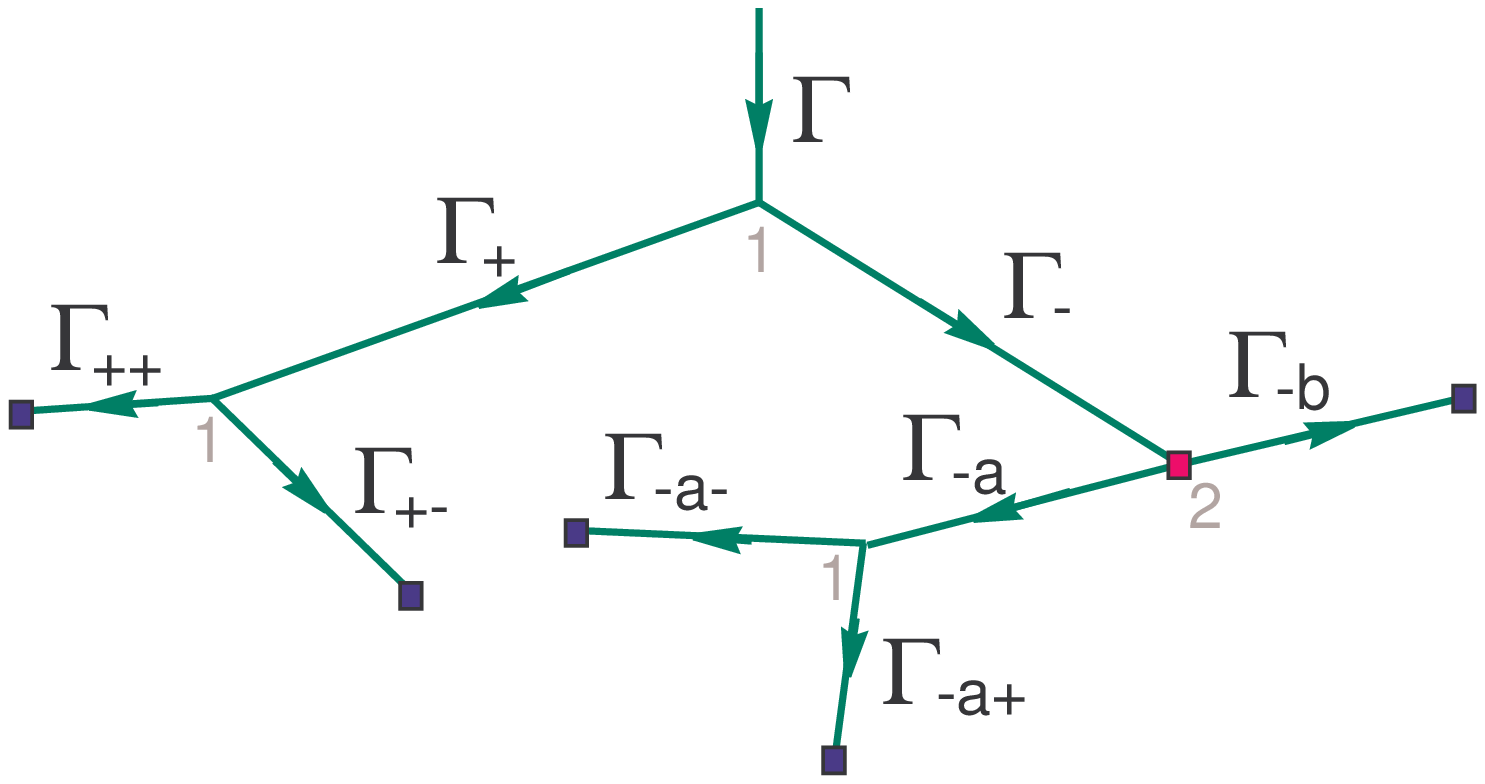,height=6.5cm}}
\caption{Example of a flow tree. The split type is indicated as a
small gray 1 or 2. The homology classes corresponding to the
different branches (and thus the branches themselves) are
recursively labeled as follows. If a branch $\Gamma_x$ splits into
two new branches through a type 1 split, the new branches are
labeled $\Gamma_{x+}$ and $\Gamma_{x-}$, where the $+/-$ is
attributed such that $\langle \Gamma_x , \Gamma_{x+} \rangle > 0$
(so $\langle \Gamma_x , \Gamma_{x-} \rangle < 0$). When the split
is of type 2, the corresponding branches are labeled
$\Gamma_{xa}$, $\Gamma_{xb}$, $\Gamma_{xc}$, and so on (in no
particular order).}\label{flowtree}}

\subsection{Stability of splits}

Type 2 splits, being fixed at an attractor point, are insensitive
to the initial moduli $z_0$, and are therefore stable for
variations of $z_0$ all over Teichm\"uller space
$\widetilde{\CM_c}$ (the covering space of moduli space $\CM_c$).

A type 1 split on the other hand is fragile. If we let $z_0$ cross
the MS wall on which the split point is located, from the side
where (\ref{prejoycecondition}) is satisfied to the side where it
is not, the split ceases to exist.

A stability condition for type 1 splits in terms of phases that
makes sense for arbitrary points $z_0$ in Teichm\"uller space
$\widetilde{\CM_c}$ can be formulated as follows. Consider the
type 1 split $\Gamma \to \Gamma_+ + \Gamma_-$, with incoming
branch starting at $z=z_0$, and take $\langle \Gamma_-,\Gamma_+
\rangle > 0$. Let $\alpha_- = \arg Z(\Gamma_-) \mod 2 \pi$ and
similarly for $\alpha_+$ and $\alpha$. The relative $2 n
\pi$-ambiguity between the phases is fixed by requiring $\alpha_-
= \alpha_+ = \alpha$ at the MS wall and taking the $\alpha_x$ to
be continuous on Teichm\"uller space. Now, since there can be at
most one solution to (\ref{tms}), we certainly need that $-\pi<
(\alpha_+ - \alpha_-)_0 < \pi$ in order for the split to exist.
Combining this with (\ref{prejoycecondition}), this yields the
condition
\begin{equation} \label{joycecondition}
 0 < (\alpha_- - \alpha_+)_0 < \pi \,
\end{equation}
(and consequently also $0 < (\alpha - \alpha_+)_0 < \pi$ and $0 <
(\alpha_- - \alpha)_0 < \pi$). This is a necessary condition.
Close to the MS wall it is also sufficient, but when moving
further away from the wall, it could fail to be so, as the split
flow might cease to exist even without negating
(\ref{joycecondition}). This will be the case if the flow is
``dragged'' through a part of the discriminant locus of $\CM_c$
and the 3-cycle vanishing there has nonzero intersection with the
3-cycle corresponding to the flow \cite{branessugra,DGR}. In
general this does not mean that the split decays like it does when
the MS line is crossed. As discussed in \cite{branessugra,DGR}, in
such situations, the original flow tree can morph into a new one
through the so-called branch creation mechanism: a new branch,
corresponding to the vanishing cycle and ending on the
discriminant locus, is ``pulled'' into existence, thus saving the
flow tree from collapse.

\section{Attractor flows as Hamiltonian (de)formations of SLGs}\label{sec:micrattr}
\setcounter{equation}{0}

\subsection{SLG deformations along attractor flows}
\label{sec:SLGhamdef}

We now turn to the stability of SLG manifolds under small
deformations of the complex structure of $X$ induced by moving
along an attractor flow. From theorem 2.14 in \cite{joyce} and the
fact that the K\"ahler class remains constant, one indeed expects
the existence of a special Lagrangian near the original one, at
least if the SLG does not degenerate during the deformation
process. In \cite{Th}, it was furthermore shown that such
deformations can be taken to be Hamiltonian.

Let us make this explicit, along the lines of section
\ref{sec:SLGdeform}. The only difference with that section is the
fact that the factor $\im(e^{-i \alpha} \Omn)$ appearing in
equation (\ref{slgflow2}) will now also have an explicit
$t$-dependence, due to the variation of the complex structure on
$X$ as given by (\ref{integrated}). To make this precise, we have
to specify a relation between $\mu$ and $t$. A convenient choice
away from the attractor point is $d t \equiv - (2 |Z| \mu)^{-1}
d\mu$. The $t$-dependence of $\im(e^{-i \alpha} \Omn)$ is then
obtained from (\ref{integrated}), where $\Gamma$ should be
understood as the harmonic representative in the cohomology class
Poincar\'e dual to $[L]$, with $L$ the SLG under consideration.
Because of this extra $t$-dependence, the left hand side of
$(\ref{dslgflow2})$ is no longer zero, but equal to $F^*[dt \wedge
\partial_t \, \im(e^{-i \alpha} \Omn)]_{t=0}$, which by
$(\ref{integrated})$ and the SLG condition equals $- dt \wedge F^*
\Gamma \, |_{t=0} = \Gamma \,|_L \wedge dt$. Therefore the
condition (\ref{etclosed}) for preservation of the SLG condition
is now
\begin{eqnarray}
 \sd \theta^{(1)} &=& 0 \\
 \sd \eta^{(2)} &=& \Gamma \,|_L \, .
\end{eqnarray}
All other identities remain the same. In particular the relation
$* \theta^{(1)} = 2 k \, \eta^{(2)}$ is unchanged. A particular
solution is therefore obtained by putting $\theta^{(1)} = 2k \,
\sd H$, $\eta^{(2)} = * \sd H$, with $H$ the up to a constant
unique solution of
\begin{equation} \label{Heq}
 \Delta H = \Gamma \,|_L \, ,
\end{equation}
with $\Delta \equiv \sd * \sd$. Such a deformation is Hamiltonian,
with Hamiltonian function $H$. Finding $H$ is equivalent to
finding the electrostatic potential on $L$ for a given charge
density $\Gamma \,|_L$. As a consistency check, observe that the
integrability condition for this problem is trivially satisfied,
as the total ``charge'' $\int_L \Gamma \,|_L = \langle
\Gamma,\Gamma \rangle = 0$. Another physical interpretation, which
is particularly useful to keep in mind for intuition in what
follows, is viewing $H$ as the equilibrium temperature on an ideal
heat conductor $L$ with heating/cooling sources given by $\Gamma
\,|_L$.

Note that because of the SLG condition and equation
(\ref{integrated}), at a regular attractor point, $\Gamma \, |_L =
0$ and therefore the solution of $(\ref{Heq})$ is trivially a
constant and the corresponding first order deformation (in $t$) of
$L$ zero. This is also required by consistency, since all
attractor flows stop at that point.

\subsection{Simple degeneration} \label{sec:simpledeg}

\FIGURE[t]{\centerline{\epsfig{file=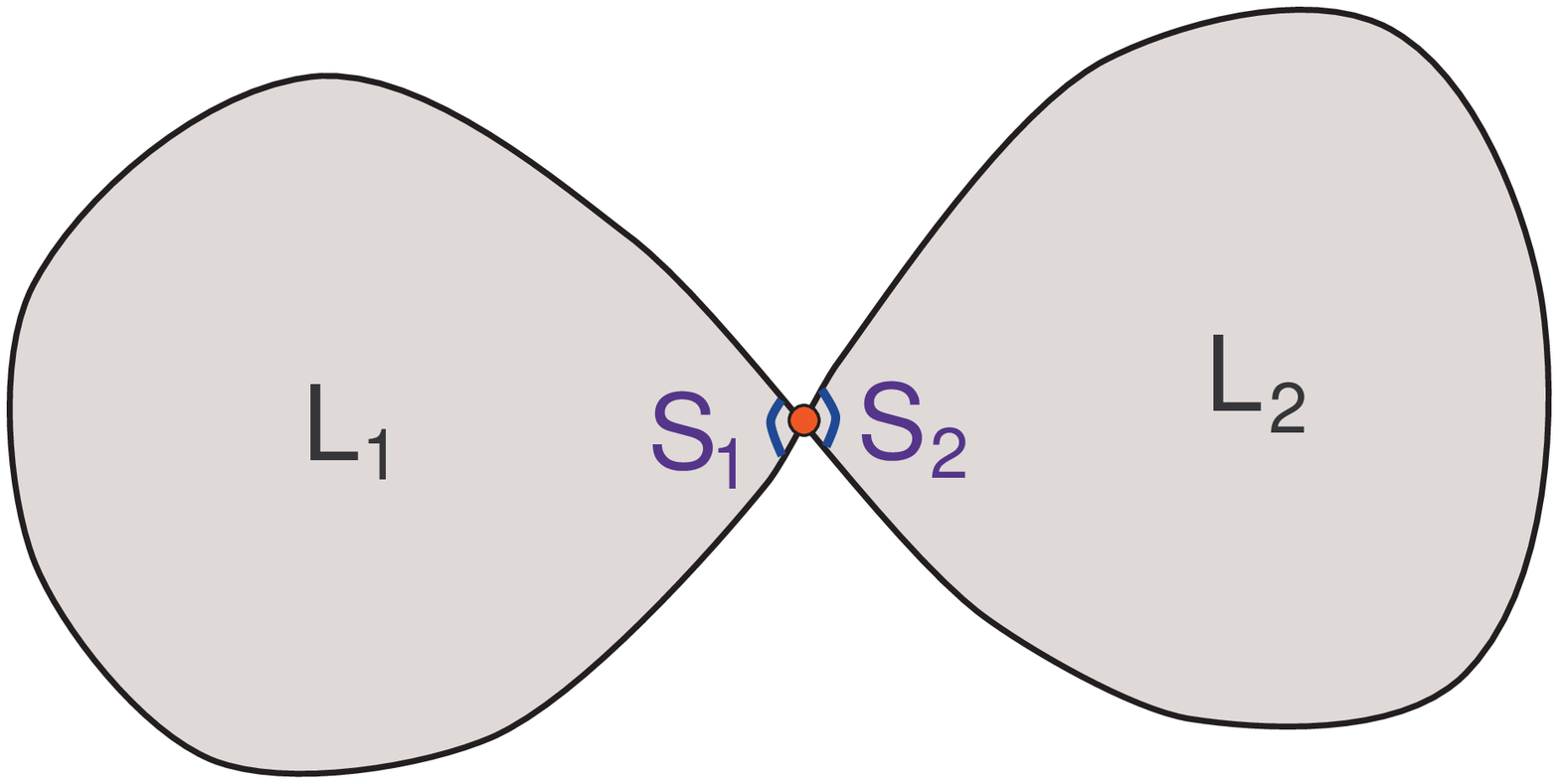,height=5cm}}
\caption{A degenerate SLG, equal to the union of two SLGs, $L_1$
and $L_2$, with equal phases and a single transversal intersection
point.}\label{glue}}

Now let us see what we get when the SLG submanifold $L$ happens to
be degenerate, equal to the union of two (smooth) SLG submanifolds
$L_1$ and $L_2$ (necessarily with equal phases). We first consider
the case where $L_1$ and $L_2$ have a single, transversal
intersection point (so we are at a point in moduli space where the
attractor flow crosses a line of $(L_1,L_2)$ marginal stability,
i.e.\ where a flow split of type 1 can occur). Without loss of
generality, we can assume that\footnote{As in other places in this
paper, when confusion is not possible, we take the liberty to use
the same notation for the SLGs and their corresponding homology
classes and their Poincar\'e dual cohomology classes.}
\begin{equation} \label{intersectchoice}
 \langle L_2,L_1 \rangle = +1.
\end{equation}
Equation (\ref{Heq}) does not have a solution on $L_1$ and $L_2$
separately, since $\int_{L_1} \Gamma|_L = \langle L_1,L \rangle =
\langle L_1,L_2 \rangle = -1 \neq 0$. To get a solution, we have
to ``connect'' $L_1$ and $L_2$ in a certain way through their
intersection point. This goes as follows. Let $S_1$ ($S_2$) be an
infinitesimally small sphere in $L_1$ ($L_2$), centered around the
intersection point. We can then consider $L=L_1 \cup L_2$ to be
the variety obtained by deleting the insides of the infinitesimal
spheres in $L_1$ and $L_2$ and ``gluing'' the two remainders
together along the spheres. With suitable orientations for $S_1$
and $S_2$, we can write $S_1 = \partial L_2$ and $S_2 =
\partial L_1$ (see fig.\ \ref{glue}). In more mathematical language this
procedure is referred to as taking the connect sum of $L_1$ and
$L_2$.

After connecting $L_1$ and $L_2$ in this way, an observer sitting
in $L_1$ will see the effect on (\ref{Heq}) of the presence of
$L_2$ as a $\delta$-function source at the intersection point:
\begin{equation} \label{deltasource}
 \int_{S_1} \eta^{(2)} = \int_{\partial L_2} * \sd H = \int_{L_2}
 \Delta H = \int_{L_2} \Gamma\, |_L = \langle L_2,L \rangle =
 1 \, .
\end{equation}
Therefore, choosing suitable spherical coordinates $r,\theta,\phi$
in $L_1$, centered at the intersection point, we can write, for
small $r$:
\begin{equation}
 \eta^{(2)} \approx \frac{1}{4 \pi} \sin \theta \, d\theta \wedge
 d \phi \, .
\end{equation}
Comparison with (\ref{ecomp}) then gives $v^\theta \approx 0
\approx v^\phi$ and $v^r \approx \frac{k}{4 \pi} \frac{1}{r^2}$.
Hence, from the definition of $v$, we get for the deformation
vector field on $L_1$ (putting $w \equiv 0$):
\begin{equation}
 \partial_t F_1^m \approx i \, \frac{k}{4 \pi r^2} \partial_r
 F_1^m \, . \label{defeq1}
\end{equation}
The above reasoning can be repeated for the deformation vector
field on $L_2$, giving only a sign difference in the final result
(because the $\eta^{(2)}$-flux through $S_2$ is opposite to the
flux through $S_1$).
\begin{equation}
 \partial_t F_2^m \approx - i \, \frac{k}{4 \pi r^2} \partial_r
 F_2^m \, . \label{defeq2}
\end{equation}
For a deformation corresponding to a change of complex structure
induced by following the attractor flow in the \emph{opposite}
direction of the flow, these two equations each acquire an extra
minus sign. From the physics of decay at marginal stability and
the work of Joyce \cite{joyce}, one expects that following the
flow in one direction will make $L_1$ and $L_2$ join into one
smooth SLG, whereas following it in the other direction will
produce a decay (into two separate SLGs). In other words, we only
expect a consistent fusion of the two SLGs for one direction.

\FIGURE[t]{\centerline{\epsfig{file=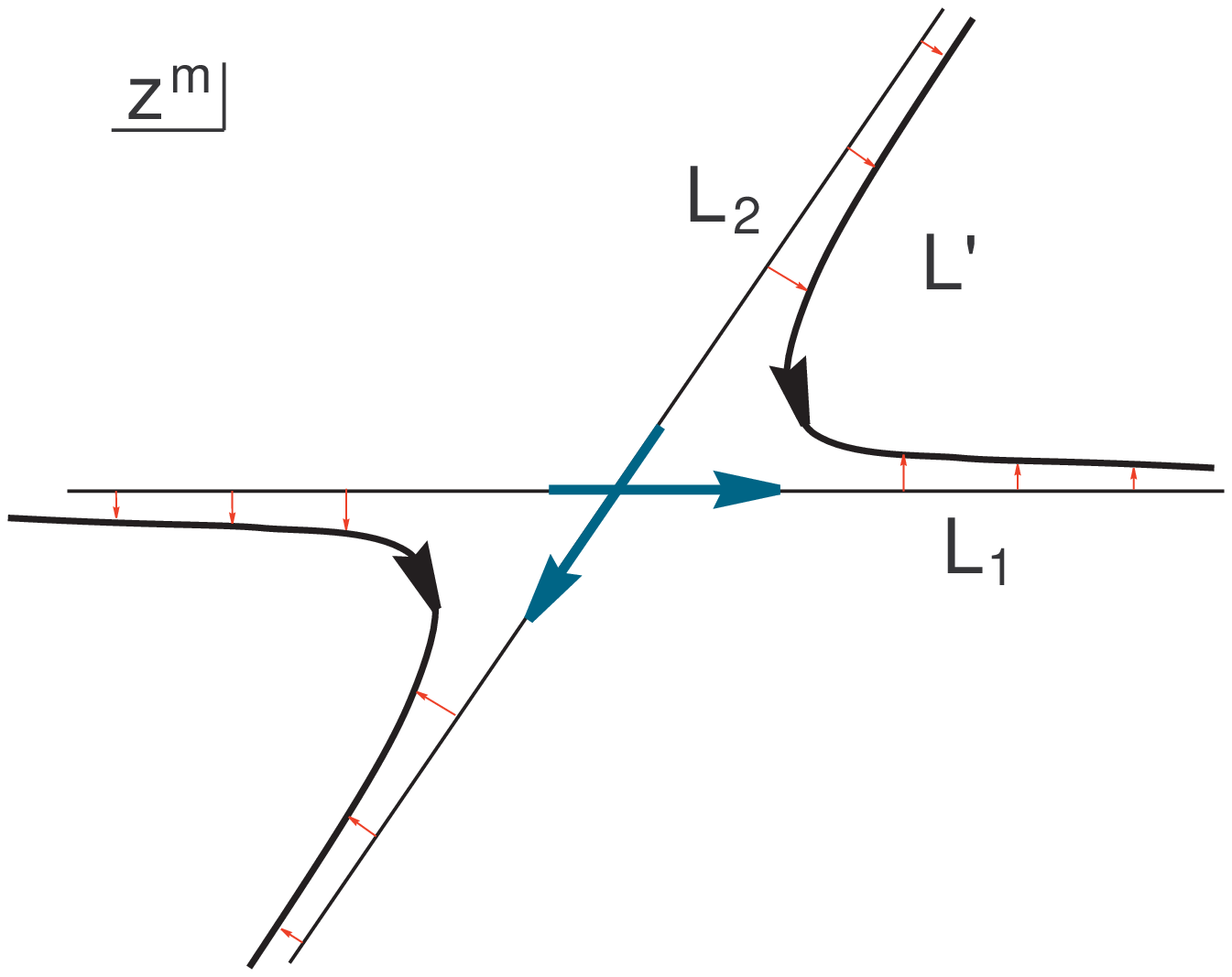,height=5.5cm}}
\caption{SLG deformation of $L_1 \cup L_2$ to $L'$ corresponding
to a change in complex structure in the direction of the attractor
flow, \emph{assuming} this deformation exist. Here the
intersection curves with the $z^m$ coordinate plane are shown,
near the intersection point $z=0$. The deformation vectors are
given by the thin red arrows. Note that, since a small but finite
deformation is shown here, the expressions
(\ref{defeq1})-(\ref{defeq2}) for infinitesimal deformations are
only valid up to a small but nonzero distance from the
intersection point. }\label{deformfig}}

\FIGURE[t]{\centerline{\epsfig{file=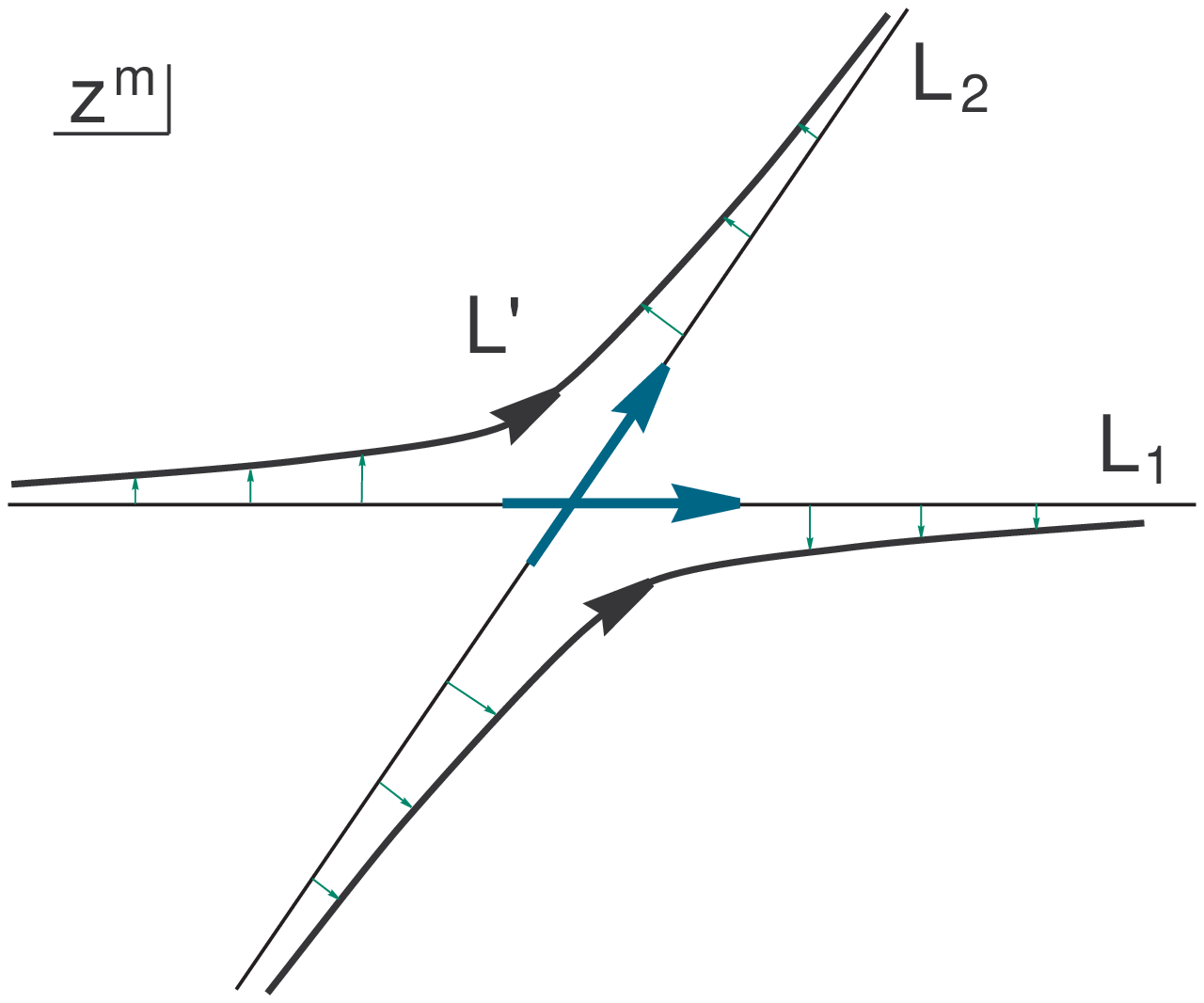,height=5.5cm}}
\caption{Same as fig.\ \ref{deformfig}, but for a change of
complex structure opposite to the attractor flow. This deformation
requires $\langle L_2,L_1 \rangle = +1$, whereas the one of fig.\
\ref{deformfig} requires the opposite sign. Only the $+1$ case is
consistent with the starting assumptions.}\label{deformfig2}}

Figuring out which one is a rather subtle business. Fig.\
\ref{deformfig} shows the intersection of a $z^m$ coordinate plane
with $L_1$ and $L_2$, tracking out certain rays (out of the
intersection point $z=0$) in $L_1$ and $L_2$, together with their
deformations according to (\ref{defeq1})-(\ref{defeq2}),
\emph{assuming} they do indeed fuse into a smooth SLG $L'$. As
explained in detail in appendix \ref{app:rays}, we get equally
oriented bases of the tangent spaces to $L_1$ and $L_2$ at $z=0$
by orienting the deformed smooth curves in each coordinate plane
$m=1,2,3$, constructing the tangent vectors to the asymptotes to
these curves in $L_1$ and $L_2$, and translating those vectors
along the original rays in $L_1$ and $L_2$ to the intersection
point, as indicated in the figure. The orientation of $L_1$ and
$L_2$ determine in turn the sign of the intersection product
$\langle L_1,L_2 \rangle$, which, using the rules outlined in
appendix \ref{app:int}, can be read off from the picture (repeated
for $m=1,2,3$): $\langle L_1,L_2 \rangle = - (-1)^3 = +1$. But
this is in contradiction with equation (\ref{intersectchoice}).
Therefore, in the direction of the attractor flow, $L_1$ and $L_2$
can \emph{not} fuse into a new SLG $L$ --- in physical terms, the
brane configuration ceases to be realizable as a BPS state.

If on the other hand we follow the attractor flow in the opposite
direction, we get the situation of fig.\ \ref{deformfig2},
yielding $\langle L_1,L_2 \rangle = - (+1)^3 = -1$, which is
consistent. We thus arrive at the important conclusion that, for
the degeneration at hand, \emph{the decay happens in the direction
of the attractor flow}. This also implies that for such a
degeneration, the stable side of the line of ($L_1$,$L_2$)
marginal stability is the one that satisfies
(\ref{joycecondition}). We thus reproduce (and extend) the SLG
stability criterion as obtained by Joyce in \cite{joyce} from an
explicit local model of the SLG degeneration.

Finally, note that $b^1(L) = b^1(L_1) + b^1(L_2)$, so the
dimension of the D3-brane moduli space of the single brane $L$
equals the sum of the dimensions of the moduli spaces of the
constituent branes $L_1$ and $L_2$.

\subsection{Degeneration with more than one transversal intersection point}
\label{sec:more than one}

\FIGURE[t]{\centerline{\epsfig{file=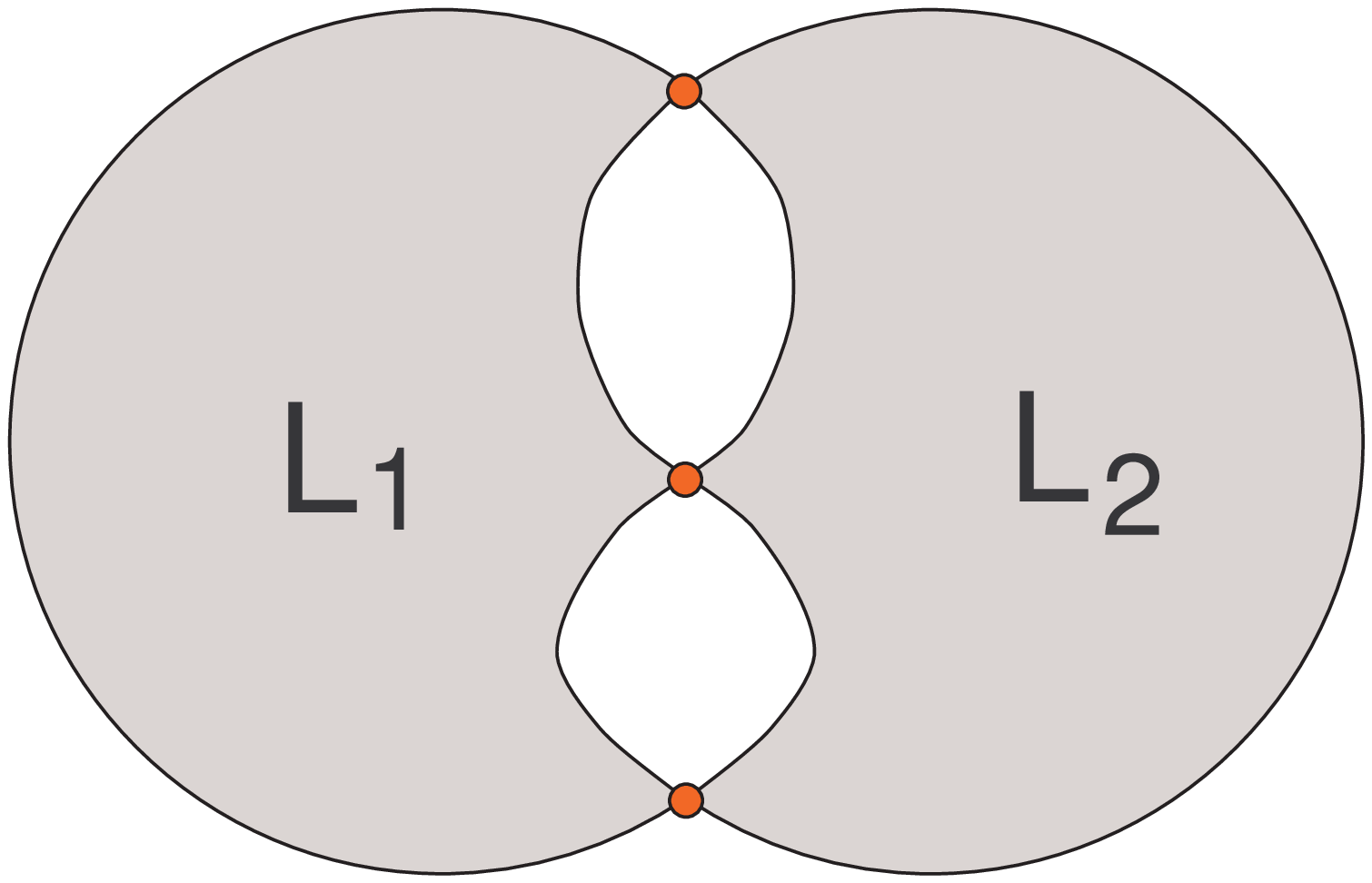,height=5cm}}
\caption{A degenerate SLG, equal to the union of two SLGs, $L_1$
and $L_2$, with equal phases and three transversal intersection
points.}\label{threepoints}}

We now turn to the case where $L_1$ and $L_2$ have multiple
transversal intersection points, say $n$, as in fig.\
\ref{threepoints}, where $n=3$. We can again assume that $\kappa
\equiv \langle L_2,L_1 \rangle  > 0$. Of the $n$ intersection
points, $n_+$ will contribute positively to $\langle L_1,L_2
\rangle$, and $n_-$ will contribute negatively. So $n=n_+ + n_-$
and $\kappa = n_+ - n_-$.

We can repeat the previous analysis, but we have a little more
freedom here: a priori we can choose along which of the
intersection points we try to glue together $L_1$ and $L_2$, and
which intersection points remain just (self-)intersection points.
We can implement this freedom by allowing but not forcing the
intersection points to appear as delta-function sources. More
concretely, the difference with the $n=1$ case is that we now have
infinitesimal spheres $S_1^{(s)}$ and $S_2^{(s)}$ for each
intersection point $P_s$, $s=1,\ldots,n$, leading to $\partial L_2
= S_1^{(1)} \cup \cdots \cup S_1^{(n)}$ and similarly for $L_1$.
Therefore, (\ref{deltasource}) now becomes
\begin{equation} \label{deltasources}
 \sum_{s=1}^n \int_{S^{(s)}_1} \eta^{(2)} = \kappa.
\end{equation}
An observer sitting in $L_1$ will still see the points $P_s$ that
get connected to $L_2$ as $\delta$-function sources, but we face
an ambiguity now: only the \emph{sum} of the corresponding charges
$Q_s$ is constrained by (\ref{deltasources}). We find for the
analog of (\ref{defeq1}), close to intersection point $P_s$:
\begin{equation}
 \partial_t F_1^m \approx \sigma \, i \, Q_s \, \frac{k}{4 \pi r^2} \partial_r
 F_1^m \, , \label{multidefeq1}
\end{equation}
with
\begin{equation} \label{Qsconstr}
 Q_s \equiv \int_{S^{(s)}_1} \eta^{(2)}
 \quad ; \qquad
 \sum_{s=1}^n Q_s = \kappa \, ,
\end{equation}
and $\sigma = +1$ in the direction of the flow, $\sigma=-1$ in the
opposite direction. Points $P_s$ through which $L_1$ and $L_2$ do
not get connected have $Q_s=0$. For the deformation vector field
on $L_2$, we get the same formulas with the opposite sign.

\FIGURE[t]{\centerline{\epsfig{file=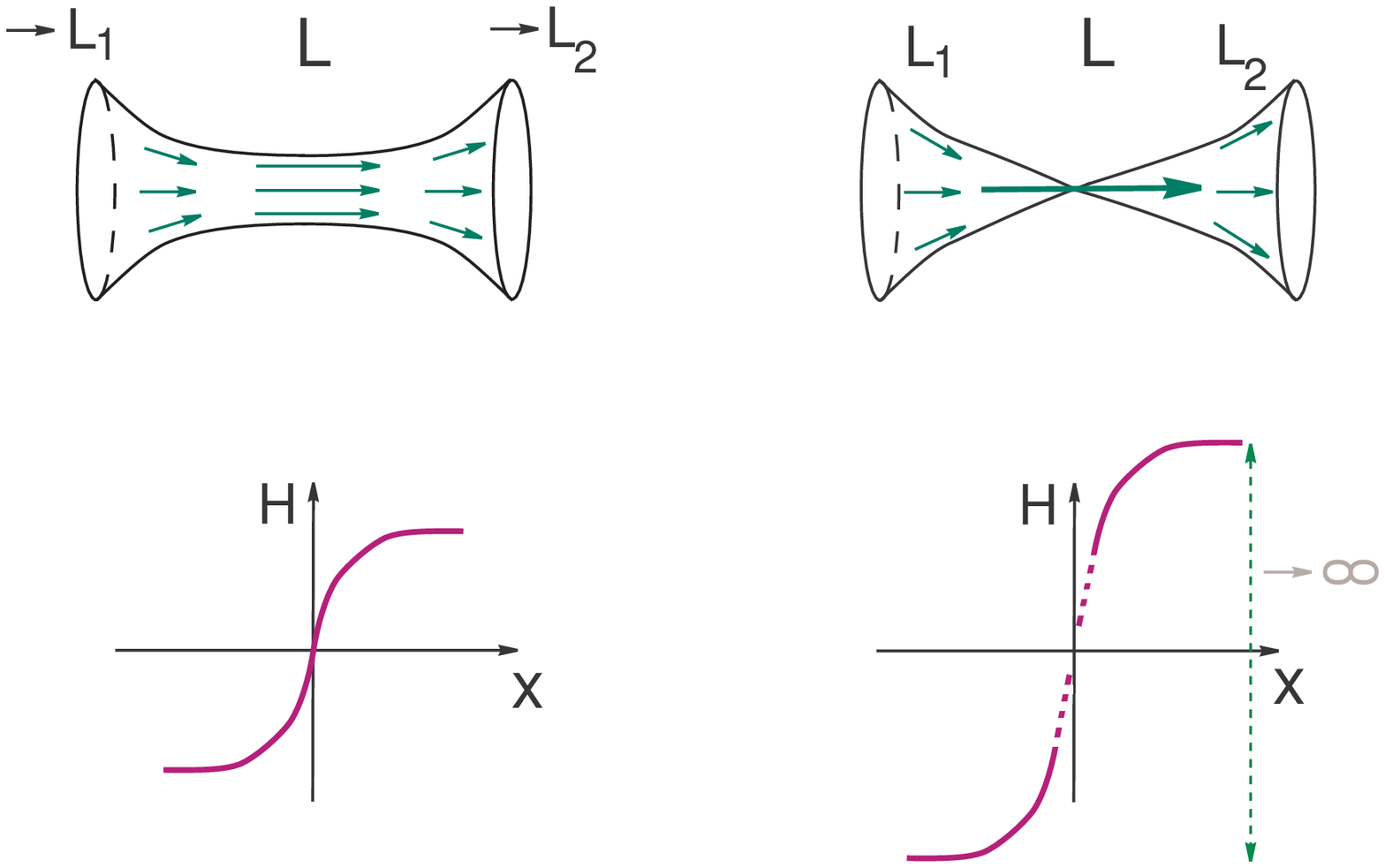,height=7cm}}
\caption{Degeneration of $L$ near $P_s$ for $\sigma Q_s < 0$. The
arrows show the ``electric field'' (or ``heat flow'') $\nabla H$,
which near degeneration (assuming the ``charge density'' $\Gamma
\, |_L$ is nonsingular near $P_s$) has approximately equal flux
through small spheres slicing the throat. The graphs on the lower
row show how the corresponding difference in $H$ between the $L_1$
side and the $L_2$ side diverges when the degeneration is
approached.}\label{degH}}

There are additional consistency constraints on the $Q_s$. First,
we will argue that in order to produce a smooth SLG $L$ from $L_1
\cup L_2$, all nonzero $Q_s$ must have the same sign. To prove
this, let us consider the opposite process, namely letting a
smooth SLG $L$ degenerate to $L_1 \cup L_2$. As long as $L$ is
smooth, the function $H$ will be a well defined, finite function
on $L$. However, when we approach the degeneration, near an
intersection point $P_s$, $H$ will approach $- \sigma Q_s/4\pi r +
\mbox{const.}$ on the $L_1$ side and $\sigma Q_s/4\pi r +
\mbox{const.}'$ on the $L_2$ side. So the difference in $H$
between the $L_1$ side and the $L_2$ side of the ``throat''
diverges to $\sigma Q_s \, \infty$, as shown in fig.\ \ref{degH}.
On the other hand, $H$ stays regular away from the degeneration
points (assuming $L_1$ and $L_2$ are regular), so to be
consistent, the difference in $H$ between the $L_1$ side and the
$L_2$ side must diverge in the same way over all throats, i.e.\
all to $+\infty$, or all to $-\infty$. This is the case if and
only if all nonzero $Q_s$ have the same sign, proving the first
assertion. (All this is intuitively quite clear if one recalls the
equilibrium temperature interpretation of $H$: $L_1$ and $L_2$ are
basically net cooled versus net heated parts of $L$, and the heat
flow is squeezed through the necks when approaching degeneration.)

Since $\kappa>0$, the second equation in (\ref{Qsconstr}) thus
implies that all $Q_s$ must be positive. Therefore, for all
``active'' intersection points $P_s$ (i.e.\ those with nonzero
$Q_s$), we have again the situation of fig.\ \ref{deformfig} for
$\sigma=+1$ and fig.\ \ref{deformfig2} for $\sigma=-1$, but the
resulting consistency constraint is more subtle now: the
contribution to the intersection product $\kappa=\langle L_2,L_1
\rangle$ of an active point $P_s$ must be negative for $\sigma=1$
and positive for $\sigma=-1$. In other words, if we go upstream
along the attractor flow, in order to obtain a smooth $L$, we have
to connect $L_1$ and $L_2$ along the intersections with sign equal
to the sign of the total intersection, whereas if we go
downstream, we need to connect along the intersection with the
opposite sign. Note that the first kind of intersections will
always be present, while the second might not (in which case $L$
necessarily decays downstream).

So from the second equation in (\ref{Qsconstr}), we see that the
connection process comes with $n_+ - 1$ degrees of freedom if we
move upstream, and $n_- - 1$ extra degrees of freedom (if any at
all) if we move downstream. It is easy to understand what the
origin is of these degrees of freedom: the manifold $L$ obtained
by gluing together $L_1$ and $L_2$ with $m$ ($m=n_+$ or $m=n_-$)
infinitesimal spheres cut out, has precisely $m -1$ extra
nontrivial 2-spheres as compared to those of $L_1$ and $L_2$
separately, i.e.\ $b_2(L,\IR) = b_2(L_1) + b_2(L_2) + m-1$, or
equivalently $b^1(L) = b^1(L_1) + b^1(L_2) + m-1$. By the results
of section \ref{sec:SLGdeform}, this means that the we have $m-1$
more degrees of freedom to deform $L$ as an SLG submanifold than
we have to deform $L_1$ and $L_2$ separately. One can think of
these degrees of freedom as the volume of small balls filling up
the connecting spheres $S_1^{(s)}$ in $X$,\footnote{This can be
made precise along the lines of \cite{joyce}.} parametrizing the
scale of the throats. The values of $Q_s$ then simply corresponds
to the rate of change of the size of these balls when we move in
moduli space to or from the degeneration point.

Note that since $n_+ = n_- + \kappa > n_-$, a transition $L \to
L_1 \cup L_2 \to L'$ downstream the attractor flow will always
\emph{lower} the dimension of the deformation moduli space of the
SLG (i.e.\ $b^1(L') < b^1(L)$).

Finally, the same kind of reasoning as above on the behavior of
$H$ yields that these degenerations must occur either at all
throats at the same time, degenerating $L$ to $L_1 \cup L_2$, or
not at all (because if one throat stays open, $H$ will stay finite
everywhere
--- in intuitive heat flow terms: there can be no temperature
divergence if the two parts stay smoothly connected). This implies
that the Hamiltonian deformations along attractor flows we are
studying here preserve the topology of $L$ unless $L$ degenerates
to a union $L_1 \cup L_2$.

Degenerations involving more than two decay products can be
analyzed in a similar way. However, if several identical copies of
the same constituent special Lagrangian are involved, additional
subtleties might arise, putting more constraints on the set of
possible active intersection points. We will not go into those
issues here, nor into the analysis of more complicated
degenerations, involving for example non-transversal
intersections.

\subsection{Type 2 degenerations} \label{sec:type2}

The degenerations we studied up to know can be called type 1
degenerations: they involve splitting into two SLGs with nonzero
intersection product and, by the discussion of section
\ref{sec:properties}, occur at a marginal stability point in
moduli space where a type 1 split can occur.

Type 2 splits on the other hand occur at attractor points and
involve splitting into charges with zero intersection product
(recall that all charges having zero intersection product with the
total charge have identical or opposite phases at the attractor
point, so in general one can expect quite some candidate decay
channels). Having zero intersection product does not mean that the
decay products are disjoint as SLG submanifolds: they can
intersect transversally with as many positive as negative
intersections, or they can intersect non-transversally, for
example along a circle. An example of the latter on a $T^6$ with
coordinates $x^1,\ldots,x^6$ is the D3-brane system
$x^1=x^3=x^5=0$ plus $x^1=x^4=x^6=0$. The corresponding splitting
of a smooth SLG $L$ into two SLGs $L_1$ and $L_2$ will be called a
type 2 degeneration.

The solution to (\ref{integrated}) for initial moduli $z_0 = z_* +
\delta z$ very close to the attractor point $z_*$ is
\begin{equation}
 \im(e^{-i\alpha} \Omn) \approx \im(e^{-i\alpha}
 \Omn)_* + \delta \im(e^{-i\alpha}\Omn) \, \mu \, ,
\end{equation}
where $\delta \im(e^{-i\alpha}\Omn) \equiv \im(e^{-i\alpha}\Omn)_0
- \im(e^{-i \alpha} \Omn)_*$. Identifying now $t$ with $\mu$, we
thus obtain for the analog of equation \ref{Heq}, describing
deformations induced by moving away from the attractor point along
the attractor flow specified by $z_0 = z_* + \delta z$:
\begin{equation} \label{Heqmutloc}
 \Delta H =  - \delta \im(e^{-i\alpha} \Omn) \,|_L = - \im(e^{-i\alpha} \delta z^a D_a \Omn)_* \,|_L \,.
\end{equation}
For the last equality, we used (\ref{varom1}).

We will not attempt to classify all possible type 2 degenerations
in this paper. Instead let us consider the case where $L=L_1 \cup
L_2$ is a trivial Lagrangian 2-fold fibration over a circle, with
the fibers of $L_1$ and $L_2$ intersecting transversally over a
finite number of points, as in the $T^6$ example given above.

The situation is then similar to the type 1 case, except that we
have to look at pictures like fig.\ \ref{deformfig} and fig.\
\ref{deformfig2} for \emph{two}-dimensional Lagrangians, fibered
over a circle. However, there is one important difference. In the
three-dimensional case, going from fig.\ \ref{deformfig} to fig.\
\ref{deformfig2} (that is, changing the side of the MS wall we
move to) results in a jump in the required intersection product of
$L_1$ and $L_2$, implying nonexistence of the SLG deformation at
one side of the MS line (since, of course, only one intersection
product can match). In the two-dimensional case, the corresponding
required intersection product (here of the 2d fibers) stays the
same, namely $+1$, so there is no longer reason to expect a decay
when crossing the MS line (which in this case means going through
the attractor point). This fits nicely with the physical
expectation (from supergravity \cite{branessugra}) that if the BPS
state exists, it should exist in the entire neighborhood of the
attractor point.

The number of deformation moduli of $L$ is also easily computed.
Denoting the fiber of $L$ by $l$, we have $b_2(L)=b_2(l) + 1$, and
similarly for $L_1$ and $L_2$. If the fibers $l_1$ and $l_2$ of
$L_1$ and $L_2$ have $n$ intersection points, we furthermore have
$b_1(l) = b_1(l_1) + b_1(l_2) + 2(n-1)$. Combining this, we find
$b_2(L) = b_1(l) + 1 = b_1(l_1) + b_1(l_2) + 2(n - 1) + 1 =
b_2(L_1) + b_2(L_2) + 2n - 3$. As in the type 1 case, the new
degrees of freedom can again be viewed as the volumes of certain
minimal 3-manifolds with boundary on $L$, only now their topology
will no longer be that of a ball, but rather that of a circle
times a disc, i.e.\ a solid 2-torus.

Many more kinds of degenerations could occur, but it is at this
point not clear to us how to proceed with a systematic analysis,
so we will leave it at this.

\section{(Dis)assembling Special Lagrangians}\label{sec:decomp}
\setcounter{equation}{0}

In this section we will explain how these results can be used to
get insight in the structure of arbitrary SLGs, and to obtain a
parametrization of their moduli spaces in favorable circumstances.

\subsection{Disassembling}

Let $L$ be a smooth special Lagrangian in the Calabi-Yau manifold
$X$ with complex structure given by $z=z_0$. If we vary the
complex structure along the attractor flow corresponding to the
homology class of $L$, we can keep $L$ special Lagrangian by
deforming it as detailed in section \ref{sec:SLGhamdef}, at least
as long as $L$ does not split.

When a type 1 split occurs along the flow, splitting $L$ into two
equal phase SLGs $L_1$ and $L_2$, the procedure can be repeated on
the two decay products separately. Correspondingly, the attractor
flow will bifurcate in a type 1 split, as described in section
\ref{sec:splitflows}. By iterating this procedure, we end up with
a flow tree containing only type 1 splits, with branches ending in
a number of attractor points, each associated to one of the
``constituent'' SLGs. The attractor points will be either at a
nonzero minimum of the corresponding central charge modulus $|Z|$,
or at a singular point\footnote{We define singular points of
moduli space as points where the volume of a 3-cycle vanishes with
respect to the square root of the volume of the entire Calabi-Yau.
Physically, those are the points where the corresponding particle
masses vanish in 4d Planck units.} with the corresponding 3-cycle
being a vanishing cycle (i.e.\ zero volume with respect to
$\sqrt{\mbox{Vol}(X)}$). In particular, it is not possible to end
up at a regular point with vanishing $Z$, since for special
Lagrangians, equation (\ref{VolSLG}) holds, so if $Z$ vanishes,
the SLG must be a vanishing cycle and we are by definition at a
singular point of moduli space.

If at an attractor point with nonzero $Z$, a type 2 degeneration
occurs, we can again split the flow, as a type 2 split, and
continue to deform the constituents along the new branches, and so
on. The whole procedure ends with a flow tree like fig.\
\ref{flowtree}, with branches ending on a set of connected special
Lagrangians, at their respective attractor points, that cannot
further be decomposed. We will call such SLGs \emph{simple}.
Clearly, any vanishing cycle at its vanishing point is simple (the
opposite is probably not true). Note that one can expect the
procedure to end after a finite number of splits, because the
volumes $|Z|$ of the SLGs decrease monotonically along the
attractor flows, and splits always result in constituents lighter
than the original SLG. For the same reason, one expects more
complicated trees to appear for higher volume SLGs (relative to
the CY volume).

In this way, we have constructed a well defined decomposition of
any special Lagrangian at a given point in moduli space into
simple special Lagrangians. As formulated, the decomposition is
unique. However, other and equally natural decompositions, based
on the same kind of Hamiltonian deformations, can be possible,
namely in those cases where at a type 1 split, we also have the
option to continue to deform along the original flow. As explained
in section \ref{sec:more than one}, this can occur for SLGs
transversally intersecting in a number points with both positive
and negative contributions to the intersection product. Thus one
SLG $L$ can correspond to several different natural flow trees.

Even if we fix this ambiguity, e.g.\ by only splitting when no
other options are open, two SLGs in the same homology class do not
necessarily give rise to the same flow tree (examples of different
flow trees within one homology class and at fixed $z_0$ can be
found in \cite{DGR}). However, generically, one can expect the
decomposition to be stable under small variations of the complex
structure moduli of $X$ and the deformation moduli of $L$, where
``stable'' means that the flow tree and the final constituents are
at most a bit (continuously) deformed. For bigger variations, this
is not necessary the case. First, there are some ``mild'' changes
possible in the topology of the tree, which can still be regarded
as continuous, like changes in connections between the different
branches, or, in the presence of a discriminant locus,
creation/annihilation of branches ending on that locus. Details
and examples of these phenomena can be found in \cite{DGR}.

A more drastic change occurs when the flow tree begins with a type
1 split and $z_0$ passes through the corresponding MS line,
causing the split and hence the entire flow tree to decay (cf.\
section \ref{sec:splitflows}). If no alternative continuation of
$L$ exists at this split point (that is, if all intersection
points contribute with the same sign to the intersection product,
see section \ref{sec:more than one}), this also implies the decay
of the associated SLG. Otherwise, a new flow tree should be
constructed, and the dimension of the moduli space of the SLG
decreases.

The map from the deformation moduli space of $L$ (at fixed $z_0$)
to the product of the deformation moduli spaces of the constituent
SLGs (at their attractor points) will in general not be injective.
Two different SLGs $L$ can end up having exactly the same
decomposition, because for instance at a type 1 split with
intersection product greater than one, some deformation degrees of
freedom disappear, corresponding to the choice of $Q_s$ when going
in the opposite direction, as explained in section \ref{sec:more
than one}. If we add at every split the relevant data resolving
this ambiguity (i.e.\ the $Q_s$ for type 1 splits) to the flow
tree data, the map to this ``dressed'' set of flow trees will be
injective. The question whether the map is also surjective will be
addressed in the next section.

The fact that every SLG has this kind of decomposition again fits
beautifully with supergravity results: the existence of such a
flow tree, with none of the branches ending on a regular zero of
$Z$, is precisely what is needed for a BPS solution to exist in
the 4d supergravity theory, as explained in \cite{branessugra}.
And if the existence of a special Lagrangian is equivalent to the
existence of a corresponding BPS state in the full string theory,
this is clearly a requirement for the consistency of the theory.
It is therefore quite pleasing to see this result appearing here.

\subsection{Assembling}

Given a certain flow tree with none of the branches ending in a
regular zero, we can try to reverse the above process to end up
with a special Lagrangian $L$ at $z_0$.\footnote{Recall that, as
shown in section \ref{sec:SLGhamdef}, deformations ``upstream''
the attractor flow do never lead to forced decays.} To do so, we
first have to pick SLGs in the homology classes specified by the
flow branches terminating in attractor points. This, of course,
requires such SLGs to exist, which is already one place where our
attempt can fail. Next we deform those SLGs upstream along the
attractor flows, as described in section \ref{sec:SLGdeform}. At
the split points, we have to fuse our SLGs together. This not
always possible, as the SLGs have to be ``connectable''. For
instance if the SLGs are disjoint, there is obviously no way to
glue them together (a simple example where this is not possible is
$X=T^6$ with $L_1$ given by $x^1=x^3=x^5=0$ and $x^1=a, x^2=x^4=0$
when $a \neq 0$). Furthermore, the subtleties that can arise in
case the splits involve multiple copies of the same SLG
constituents, briefly mentioned at the end of section
\ref{sec:more than one}, can also produce obstructions to
successful gluing.\footnote{This might very well be the key to the
resolution in this context of the ``s-rule problem'' discussed in
\cite{DGR,argyres}.} So there will be restrictions on the set of
SLGs we start with in order to be able to complete the job. In
particular, this implies that the map between SLGs $L$ and
candidate constituents for a given flow tree will not be
surjective in general. Figuring out these restrictions could be a
difficult task in general, though it seems doable for simple
models, like $X=T^6$.

In conclusion, if the issue of restrictions on the constituents
can be dealt with, this (de)composition of SLGs provides a
classification scheme for SLGs in Calabi-Yau threefolds, and in
favorable circumstances a parametrization of their moduli spaces:
if the moduli spaces of the constituents are known, the moduli
space of the assembled Lagrangian is obtained by combining the
constituent moduli spaces, adding the extra ``$Q_s$'' degrees of
freedom at the split points, and making some identifications if
necessary. A big advantage of this setup is that it is not
necessary to construct the SLGs explicitly.

\subsection{Comparison with $\Pi$-stability}

Though our basic stability criterion for SLGs assembled in this
way is simply that the corresponding flow tree must exist for the
given $z_0$, it is useful to compare this with the (extended)
$\Pi$-stability conjecture of Douglas et al.\
\cite{DFR,Dcat,Dstrings}. This criterion roughly states that a
``topological'' ``object'' $C$ is physically stable if for any two
stable objects $A$ and $B$ of which $C$ can be considered to be
``made of'', the ``triangle'' $A \to C \to B$ is stable, the
latter meaning that the ``morphism grades'' between two subsequent
objects involved all lie between 0 and 1. For a precise definition
of the words between quotation marks, we refer to
\cite{Dcat,Dstrings}. Essentially the objects here are graded
special Lagrangians, the grade distinguishing between different
possible $\IR$-valued phases of the SLG. The morphism grade
between two subsequent entries in the triangle given above is
simply their (moduli dependent) phase difference divided by $\pi$.
Thus, given the set of stable objects at one point in moduli
space, this stability criterion yields in principle the stable
objects at all other points.

A similar but not identical statement holds here, essentially
because of the stability condition (\ref{joycecondition}) for type
1 splits.\footnote{Recall that type 2 splits cannot be
destabilized by variations of $z_0$, so as far as the moduli
dependence of stability is concerned, we need only to consider
type 1 splits.} If we formally associate to a type 1 split $L \to
L_+ + L_-$ (in the notation of fig.\ \ref{flowtree}) the triangle
$L_+ \to L \to L_-$, we get indeed from (\ref{joycecondition})
that in order for this split to exist, we need the grades
$(\alpha-\alpha_+)_0/\pi$ and $(\alpha_- -\alpha)_0/\pi$ to be
between 0 and 1, or in the terminology of \cite{Dcat,Dstrings},
that the triangle $L_+ \to L \to L_-$ is stable (at $z_0$).

Obviously, though the setup is quite different, this is formally
similar to the $\Pi$-stability criterion. However, even after
making natural identifications between the two setups, there are
differences. First, in the $\Pi$-stability criterion, to verify
the stability of $C$, the objects $A$ and $B$ in the triangles $A
\to C \to B$ to be checked must be stable \emph{at $z_0$}. This is
not necessarily the case in our setup: we need that $A$ and $B$
are stable \emph{at the split point $z_s$} (to verify their
stability there, in the case that those branches again have type 1
splits, the above analysis has to repeated, but with $z_0$
replaced by $z_s$, and so on if more type 1 splits occur). If
$z_0$ is not too far from $z_s$, this will be equivalent, but not
necessarily in general. On the other hand, to check
$\Pi$-stability, one might have to verify several potentially
destabilizing triangles involving $C$, all at $z_0$ (presumably
corresponding to the possible occurrence of more than two
constituents in our flow trees). In our case, there is only one
triangle to check at $z_0$, the one corresponding to the first
split (assumed to be of type 1). However, in general, there
\emph{are} other, ``lower level'', triangles to check, namely
those corresponding to the subsequent splits, but a priori they
should be verified at subsequent split points, not at $z_0$.

Since the differences are of the apples versus oranges kind, it is
not inconceivable that they ``annihilate'' each other up to a
certain extent, and that in many or even all cases the two
criteria are actually equivalent. To settle this, one should try
to find specific examples separating the two (or prove
equivalence, of course), but we will leave this for future work.

\section{Some (simple) examples} \label{sec:examples}

In this section we will illustrate our results with some simple
examples, mainly on $T^6$. A much more elaborate study of this
case will appear elsewhere \cite{BD}.

\subsection{Type 1: the mirror of D0-D6 on the diagonal $T^6$}

\FIGURE[t]{\centerline{\epsfig{file=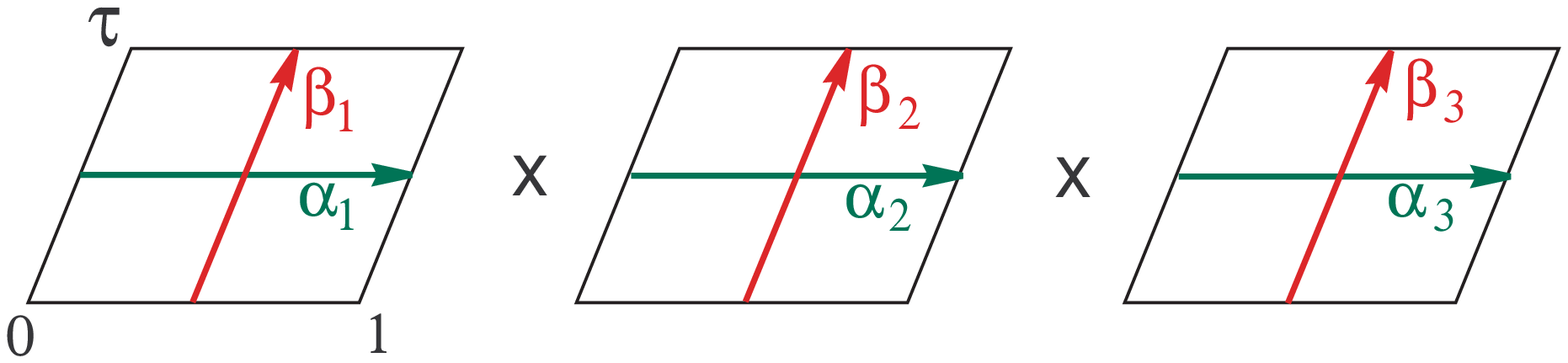,height=4cm}}
\caption{The diagonal torus with modulus $\tau$. The 3-brane
mirror to the $D0$ is $\alpha_1 \times \alpha_2 \times \alpha_3$,
the one mirror to the $D6$ is $- \beta_1 \times \beta_2 \times
\beta_3$.}\label{diagtorus}}

Let $X_\tau$ be the diagonal $T^6$ \cite{M} with modulus $\tau$,
that is, $X_\tau=E_\tau \times E_\tau \times E_\tau$, with
$E_\tau$ the 2-torus with standard complex structure parameter
$\tau$ (valued in the upper half plane), as shown in fig.\
\ref{diagtorus}. For $D3$-brane charges respecting the permutation
symmetry of the three 2-tori, this subfamily $\{X_\tau\}_\tau$ of
6-tori is closed under the attractor flow equations, making this a
particularly simple example.

Type IIB string theory on $X_\tau$ is mirror (or
T-dual\footnote{By T-dualizing along the three $\alpha_i$-cycles})
to IIA on $Y=E'_\tau \times E'_\tau \times E'_\tau$, with
$E'_\tau$ the 2-torus with area $\im \tau/2$ and $B$-field flux
$\re \tau/2$ (which together determine the complexified K\"ahler
class of $Y$). The IIA $D0$- and $D6$-brane are mirror to
respectively the $\alpha_1 \times \alpha_2 \times \alpha_3$
D3-brane and the $-\beta_1 \times \beta_2 \times \beta_3$ D3-brane
(see fig.\ \ref{diagtorus}).\footnote{Our orientation conventions,
and hence what we call a brane and what an anti-brane, are such
that the period $Z(D0) \sim 1$, where ``$\sim$'' means positively
proportional, $Z(D2) \sim \tau$, $Z(D4) \sim -\tau^2$ (i.e.\ such
that D4 and D0 have equal phases at $\re \tau = 0$) and $Z(D6)
\sim -\tau^3$ (such that D6 and D2 have equal phases at $\re \tau
= 0$).} It is convenient to label also the IIB D3-branes by their
IIA names and we will do so henceforth, hoping not to cause
confusion.

The K\"ahler form on $X_\tau$ can be taken to be $\omega = dz^1
\wedge d\bar{z}^{\bar{1}} + dz^2 \wedge d\bar{z}^{\bar{2}} + dz^3
\wedge d\bar{z}^{\bar{3}}$, and the normalized holomorphic 3-form
$\Omega = (2 \im \tau)^{-3/2} \, dz^1 \wedge dz^2 \wedge dz^3$.
The trivial flat embeddings of the D3-branes under consideration
are special Lagrangian, giving a 3-parameter moduli space (with
torus topology) for each of them (complex if Wilson lines are
included).

\FIGURE[t]{\centerline{\epsfig{file=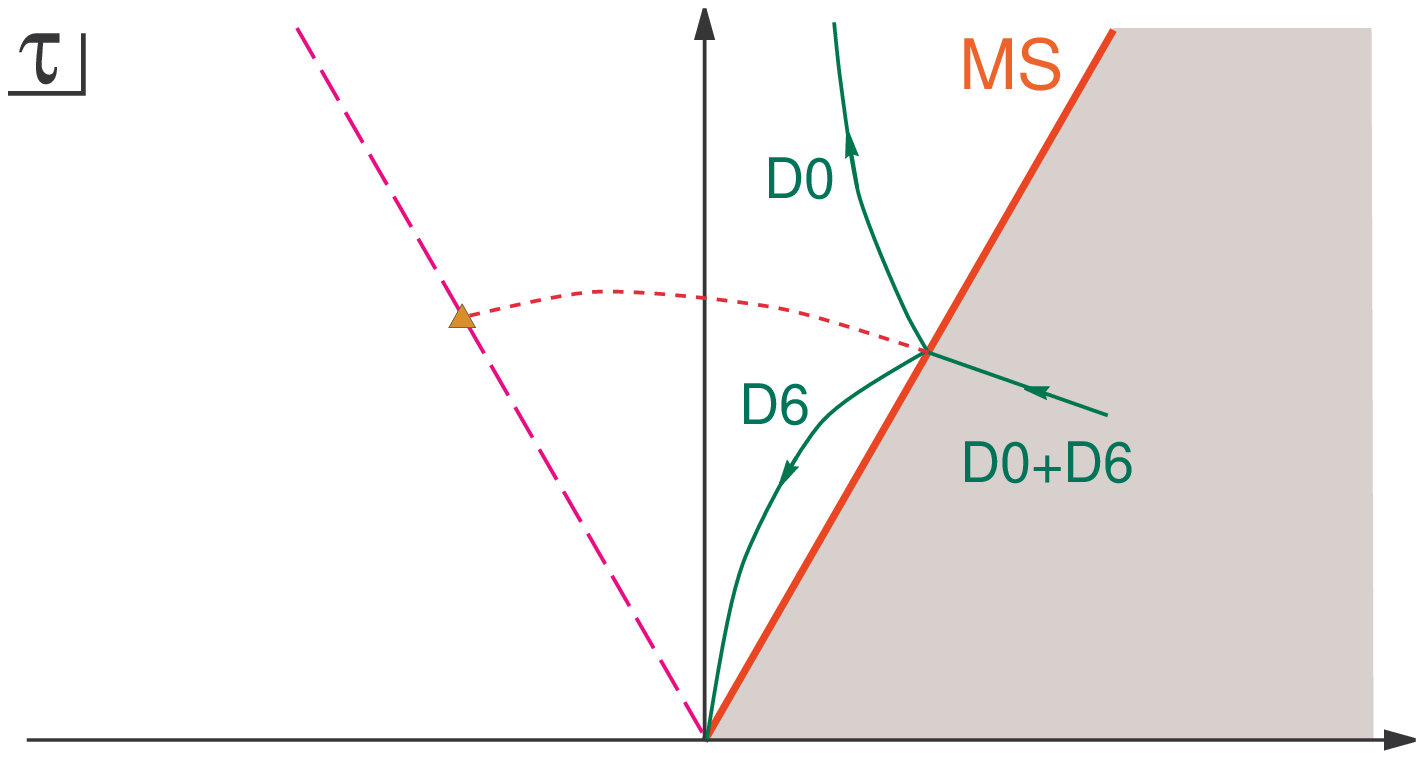,height=5cm}}
\caption{The shaded region $\im \tau < \sqrt{3} \re \tau$ shows
the stable domain of the composite (or ``bound state'') D6-D0 SLG.
A typical split attractor flow corresponding to this SLG is
sketched in green. The $D0$ and $D6$ branches have attractor
points at the ``large complex structure'' singularities
$\tau=i\infty$ and $\tau=0$ respectively. The dashed straight line
$\im \tau = - \sqrt{3} \re \tau$ is the line of $(D0,D6)$ anti-MS.
It contains a zero for the period $Z(D6+D0)$, at $\tau=e^{2
i\pi/3}$. The would-be single flow for $D6+D0$ (the dotted red
curve) crashes at this zero.}\label{D0D6}}

The D6 and the D0 have one transversal intersection point.
Starting at a D6-D0 marginal stability line and moving into the
stable domain, they will therefore fuse into a single SLG, as
explained in section \ref{sec:simpledeg}. The quotient of the
$D6$-period and the $D0$-period is $-\tau^3$, so $D6$-$D0$
marginal stability occurs at the line $\im \tau = \sqrt{3} \re
\tau$, as shown in fig.\ \ref{D0D6}. The intersection product is
easily computed using the rules of appendix \ref{app:int}:
$\langle D0,D6 \rangle = +1$. The stability condition
(\ref{joycecondition}) thus becomes
$0<\alpha_{D0}-\alpha_{D6}<\pi$, corresponding to the shaded area
$\im \tau < \sqrt{3} \re \tau$ in the figure. In the white region,
the SLG ceases to exist: it splits in a pure D6 and a pure D0 SLG,
with different phases; physically this means the state is no
longer BPS. Going to the IIA interpretation, this agrees with the
results of \cite{park} (at marginal stablity) and \cite{WiD6D0}
(on the full moduli space), where it was shown that a
supersymmetric D6-D0 bound state on $T^6$ exists, but only for a
certain range of B-field fluxes.

The number of moduli of these composite (``bound state'') D6-D0
SLGs is 6, as is easily computed using the formula at the end of
section \ref{sec:simpledeg}. The map between the composite SLGs
and their consituents, discussed in section \ref{sec:decomp}, is
furthermore one to one here, so the deformation moduli space will
simply be the product of the moduli spaces of $D0$ and $D6$.

The deformation equation (\ref{Heq}) and its solution can also be
made more explicit here, at least on the MS line. Writing $dz^m =
du^m + \tau \, dv^m$ for $m=1,2,3$, we have $\Gamma = du^1 \wedge
du^2 \wedge du^3 + dv^1 \wedge dv^2 \wedge dv^3$ and we can take
$L_1=D6$, $L_2=D0$, $L=L_1 \cup L_2$. Then on the $D0$ part of
$L$, (\ref{Heq}) becomes
\begin{equation}
 \Delta H = \sigma \, [1 - \delta^3(\su-\su_0)] \, du^1 \wedge du^2 \wedge du^3 = \sigma \, [1 - \delta^3(\su)] \, dV \, ,
\end{equation}
and on the $D6$ part
\begin{equation}
 \Delta H = \sigma \, [1 - \delta^3(\sv-\sv_0)] \, dv^1 \wedge dv^2 \wedge dv^3 = - \sigma \, [1 - \delta^3(\sv)] \, dV \,
 ,
\end{equation}
where $dV$ is the volume element, $\su_0$ and $\sv_0$ are the
coordinates of the intersection point, and $\sigma = +1$ for a
change in $\tau$ in the direction of the attractor flow,
$\sigma=-1$ in the opposite direction. One could write down
explicit solutions to these equations, but we won't do so here.

It is possible to generalize all this to less trivial examples,
like for example replacing the $\beta_m$-cycles by $n \beta_m +
\alpha_m$, producing an intersection product equal to $n^3$. Instead
of doing that, let us consider an example involving a type 2
split.

\subsection{Type 2: the mirror of D0-D4 on the diagonal $T^6$}

The D3-brane corresponding to the type IIA D0 is still $D0 \equiv
\alpha_1 \times \alpha_2 \times \alpha_3$. The D4 is a bit more
complicated, because we require the symmetry between the three
$E_\tau$ tori to be respected. The natural IIB D3-brane system to
consider is then $D4 \equiv -(\alpha_1 \times \beta_2 \times
\beta_3 + \beta_1 \times \alpha_2 \times \beta_3 + \beta_1 \times
\beta_2 \times \alpha_3)$, which has 3 units of D4-brane charge on
the IIA side, and period $Z(D4) = -3 \, (2 \im \tau)^{-3/2} \,
\tau^2$. Its intersection product with D0 is zero.

\FIGURE[t]{\centerline{\epsfig{file=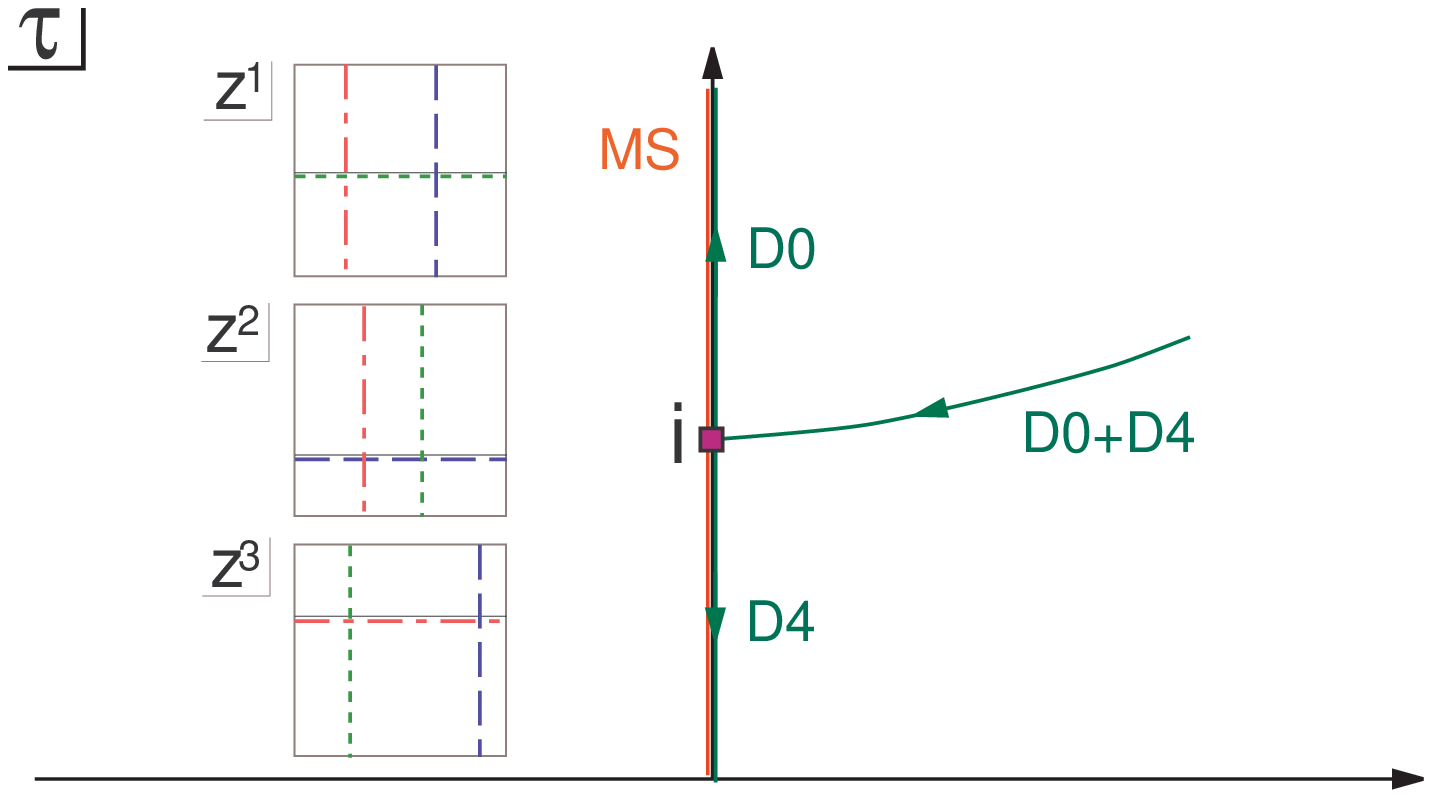,height=5.5cm}}
\caption{$\tau$-plane with sketch of a typical (type 2 split)
attractor flow for the D4-D0 composite SLG (which is everywhere
stable). The intermediate attractor point is at $\tau=i$. Also
included is a picture of a connectable D4-D0 system in $T^6$ at
this attractor point, where D4 and D0 have equal phases and hence
can suppersymmetrically coexist.}\label{D0D4}}

The homology class $D0+D4$ has period $Z(D4+D0)=(2 \im
\tau)^{-3/2} \, (1-3\tau^2)$, which has a nonzero minimal norm (=
regular attractor point) at $\tau = i$, where $Z=Z_*=\sqrt{2}$. At
the attractor point, D4 and D0 have the same phase $\alpha=0$, so
the D4-D0 system is supersymmetric there. The same holds for all
points on the imaginary $\tau$ axis.

Deformations away from $\tau=i$, fusing the separate branes into
one SLG, are governed by equation (\ref{Heqmutloc}). For this
equation to make sense (and fusing to be possible), the branes
must have common points, along which they can get glued together.
This is the case if and only if of each of the three D4
components, the $\alpha$ cycle coincides with the corresponding
$\alpha$ cycle of the D0, as sketched in fig.\ \ref{D0D4}. Then
the D0 brane intersects each of the D4 components in a circle.
Thus, any D4 component put together with the D0 brane can be seen
as a fibration over a circle, with transversally intersecting
2-dimensional SLG fibers (with intersection $+1$), and we are
essentially in the situation considered in section
\ref{sec:type2}.

To make (\ref{Heqmutloc}) explicit, denote the position of the D4
component $\alpha_1 \times \beta_2 \times \beta_3$ by
$(v^1_{(1)},u^2_{(1)},u^3_{(1)})$ and similarly for the other
three components. The requirement of coincidence then implies that
the position of the D0 is given by
$(v^1_{(1)},v^2_{(2)},v^3_{(3)})$. A simple calculation shows that
\begin{equation}
 D_\tau \Omega = i (2 \im \tau)^{-5/2} (d \bar{z}^{\bar{1}} \wedge dz^2 \wedge dz^3
  + dz^1 \wedge d \bar{z}^{\bar{2}} \wedge dz^3
  + dz^1 \wedge dz^2 \wedge d \bar{z}^{\bar{3}}) \, ,
\end{equation}
so (\ref{Heqmutloc}), for an infinitesimal deviation $\delta \tau$
from $\tau=i$, becomes on the D0 part:
\begin{eqnarray}
 \Delta H = - 2^{-5/2} \re \delta \tau
 && [ 3
 - \delta(u^2-u^2_{(1)})\delta(u^3-u^3_{(1)}) \nonumber
 - \delta(u^1-u^1_{(2)})\delta(u^3-u^3_{(2)}) \\
 &&
 - \delta(u^1-u^1_{(3)})\delta(u^2-u^2_{(3)})] \, dV \, ,
\end{eqnarray}
on the D4 component $\alpha_1 \times \beta_2 \times \beta_3$:
\begin{equation}
 \Delta H = 2^{-5/2} \re \delta \tau \,
 [1 - \delta(v^2-v^2_{(2)}) \delta(v^3-v^3_{(3)}) ] \, dV \, ,
\end{equation}
and similarly for the other two D4 components. The delta-function
sources, localized on the intersecting circles, will produce the
required fusions, yielding an SLG with 9 deformation moduli
(basically the positions of the four constituents minus 3 because
of the coincidence constraint). Note that if $\re \delta \tau =
0$, we get a constant $H$, so the branes stay at their original
$(u^m,v^m)$ positions on the $T^6$. Indeed, on the imaginary axis,
the D4-D0 system stays special Lagrangian without deformation.

Again one could consider more complicated (higher charge)
configurations and try to deduce the topology of their moduli
spaces, which upon quantization should provide for example a
microscopic computation of the corresponding black hole entropy.

\subsection{The (no longer) mysterious $|10000\rangle_B$ brane on the Quintic}

In \cite{BDLR}, using CFT techniques, a number of BPS states was
established to exist in IIA string theory at the Gepner point on
the the Quintic. One of those states, labeled $|10000\rangle_B$,
caused some confusion for a while, as it doesn't correspond to a
regular single attractor flow, which at the time was thought to
imply that it doesn't have a corresponding BPS supergravity
solution. The issue got cleared up in \cite{branessugra}, where it
was noted that it gives rise instead to a (type 1) split flow,
which in turn is associated to a multicentered BPS supergravity
solution. The two branches of this split flow correspond to SLGs
vanishing at two different copies of the conifold point in
Teichm\"uller space, with mutual intersection product equal to 5.
Thus, the results of this paper tell us that there exists indeed
an SLG in the required homology class at the Gepner point, formed
by fusing these two SLGs together. In \cite{DGR} strong numerical
evidence was presented showing that beyond the split point, no
consistent flow trees exist for this homology class, so no SLGs
either. Such forced decays occur when all intersection points have
the same sign (section \ref{sec:more than one}), so we can assume
the total number of intersection points to be 5. Therefore, since
the vanishing cycles are 3-spheres, which have $b^1=0$ and
therefore no deformation moduli, the number of deformation moduli
of the resulting SLG is, according to the formulas at the end of
section \ref{sec:more than one}, equal to $5-1=4$. This agrees
with the number of moduli found in \cite{BDLR}.

\section{Conclusions}

We have constructed a (de)composition/classification scheme for
arbitrary special Lagriangian submanifolds in a Calabi-Yau 3-fold,
based on attractor flow trees. In favorable circumstances, this
allows one to extract nontrivial information about existence,
stability domains and deformation moduli spaces of special
Lagrangians without having to construct them explicitly.
Considering the virtual impossibility to construct generic SLGs in
compact manifolds, this is an important simplification, similar to
the way dealing with Calabi-Yau spaces becomes possible without
explicit construction by invoking Yau's theorem and other results
in algebraic geometry.

Clearly this construction may have many useful applications in
string theory, and through string theory in $\N=1$ field theories,
as these are (at large volume) the world-volume theories of
D-branes filling the four-dimensional noncompact space and
wrapping an SLG. It should be noted though that stringy
corrections are expected to the dynamics and moduli spaces of such
D-branes as compared to their classical geometric counterparts,
certainly away from large volume. However, the role of the
attractor flow trees themselves will conceivably remain unchanged
also after corrections to the BPS condition on the brane
embedding. This is because the decomposition along a flow tree has
the physical interpretation of moving the BPS ``particle'', no
matter how it is represented, adiabatically from spatial infinity
into the interior of the corresponding (large N) supergravity
solution, letting it decay wherever it is forced to, and repeating
all this on the decay products (along different paths) and so on,
all the way to the respective black hole cores. Because the IIB
complex structure moduli space is exact at tree level, this
picture should remain intact after corrections. In fact, it is
possible that in general (away from large volume, or perhaps even
already at large volume) the Hamiltonian flows on SLGs as defined
in section \ref{sec:micrattr} are not well defined,\footnote{I
thank R.\ Thomas for pointing this out to me.} leading to nasty
singularities and the like, and that to have a well defined
deformation flow avoiding singularities, such that the microscopic
interpretation of the flow tree picture makes sense, stringy
corrections \emph{must} be taken into account.

Unavoidably for a paper of limited size in this context, there are
quite some loose ends. We hope we have given the reader at least
an idea of how to proceed in principle. A first step to get more
insight would be to construct a number of explicit examples, for
instance on $T^6$. From the mathematical side, an open problem is
(to my knowledge) under what conditions the existence of an SLG at
a regular attractor point (nonzero $|Z|_*$) is guaranteed. (This
is related, at large volume / large complex structure, to the
``positive discriminant'' conditions for vector bundles.) Getting
a better grip on this would increase the usefulness of attractor
flow trees for establishing existence of SLGs. From the physics
side, apart from applications to $\N=1$ gauge theories, it would
be interesting to see what quantization of SLG moduli spaces
obtained through our construction (if feasible) could teach us
about black hole entropy and other four dimensional space-time
properties, especially in the light of the rather special
(multi-centered) black hole solutions appearing in the
corresponding low energy four-dimensional supergravity theories.


\acknowledgments

I would like to thank Ben Craps and Brandon Bates for discussions,
and Richard Thomas for useful comments on the first version of
this paper.


\appendix
\setcounter{equation}{0}

\section{Connect sums, trajectories and orientations} \label{app:rays}

In section \ref{sec:simpledeg}, we introduced the connect sum of
two 3-folds $L_1$ and $L_2$ with one transversal intersection
point $P$ as the singular variety $L$ obtained by cutting out
infinitesimal spheres around the intersection point in $L_1$ and
$L_2$ and gluing the two manifolds together along the spheres. In
this appendix, we will develop a practical way to determine a
basis of tangent vectors at $P$ in $L_1$ and $L_2$ which have both
positive (or both negative) orientation, given the way $L_1$ and
$L_2$ are connected. This is needed to compute the intersection
product of $L_1$ and $L_2$ (as explained in appendix
\ref{app:int}).

For our purposes here, we can locally model $L_1$ and $L_2$ as two
positively oriented copies of $\IR^3$, and $L$ as the manifold
obtained by removing spheres of radius $\epsilon \to 0$ around the
origin and gluing the remainders together along those spheres.
More precisely, picking spherical coordinates
$(r_1,\theta_1,\phi_1)$ resp.\ $(r_2,\theta_2,\phi_2)$ in the two
copies of $\IR^3$, we make the identifications
\begin{eqnarray}
 r_1 &=& \epsilon^2 / r_2 \\
 \theta_1 &=& \theta_2 \label{thetatrans} \\
 \phi_1 &=& \pi - \phi_2 \, . \label{phitrans}
\end{eqnarray}
The identifications are chosen such that $dr_2 \wedge d\theta_2
\wedge d\phi_2$ is positively proportional to $dr_1 \wedge
d\theta_1 \wedge d\phi_1$, making the orientation of $L$ well
defined.

Now imagine three particles in our local model of $L$, flying
along the negative $x$, $y$ and $z$ axes, coming from infinity in
the part of $L$ identified with $L_2$, and moving towards the
sphere $r_2=\epsilon$. In other words, the first particle comes
from $\theta_2=\pi/2, \phi_2=\pi$ (and large $r_2$), the second
from $\theta_2=\pi/2, \phi_2=-\pi/2$, and the third from
$\theta_2=\pi$. At any time, the velocity vectors of these
particles, when further translated to the origin, form a
positively oriented basis of $\IR^3$ at the origin, thus giving a
positively oriented basis of tangent vectors of $L_1$ at $P$. Now
when the particles pass through the sphere $r_2=\epsilon$ in $L$,
they pop out in the other copy of $\IR^3$ at $r_1=\epsilon$,
moving outwards, according to (\ref{thetatrans})-(\ref{phitrans})
respectively at $\theta_1=\pi/2, \phi_1=0$, at $\theta_1=\pi/2,
\phi_1=3\pi/2$ and at $\theta_1=\pi$. When translated to the
origin of the copy of $\IR^3$ associated to $L_1$, their velocity
vectors are easily seen to be again a positively oriented basis of
$\IR^3$. (Basically the last two basis vectors flip sign with
respect to the basis obtained in the other copy of $\IR^3$.)

The upshot of all this is that if we imagine a set of three
particle trajectories through $L$ going from asymptotic $L_2$ to
asymptotic $L_1$, and we produce bases of the tangent spaces to
$L_1$ and $L_2$ at $P$ associated to the asymptotic trajectories
as outlined above, we get two identically oriented bases.

Note that this result is not as self-evident as it might seem: in
the case of \emph{two}-dimensional special Lagrangians, the
opposite is true, as can be checked by repeating the above
reasoning for two copies of $\IR^2$. The bases associated to
asymptotic trajectories now have \emph{opposite} orientations.
Essentially, this is because only \emph{one} basis vector flips
sign upon passing trough the circle connecting the $L_1$ and $L_2$
parts of $L$, resulting in a basis of opposite orientation.

In general, for odd (even) dimensional SLGs, the bases associated
to the two trajectory asymptotes will have equal (opposite)
orientations.

\section{Computing intersection products} \label{app:int}

\FIGURE[t]{\centerline{\epsfig{file=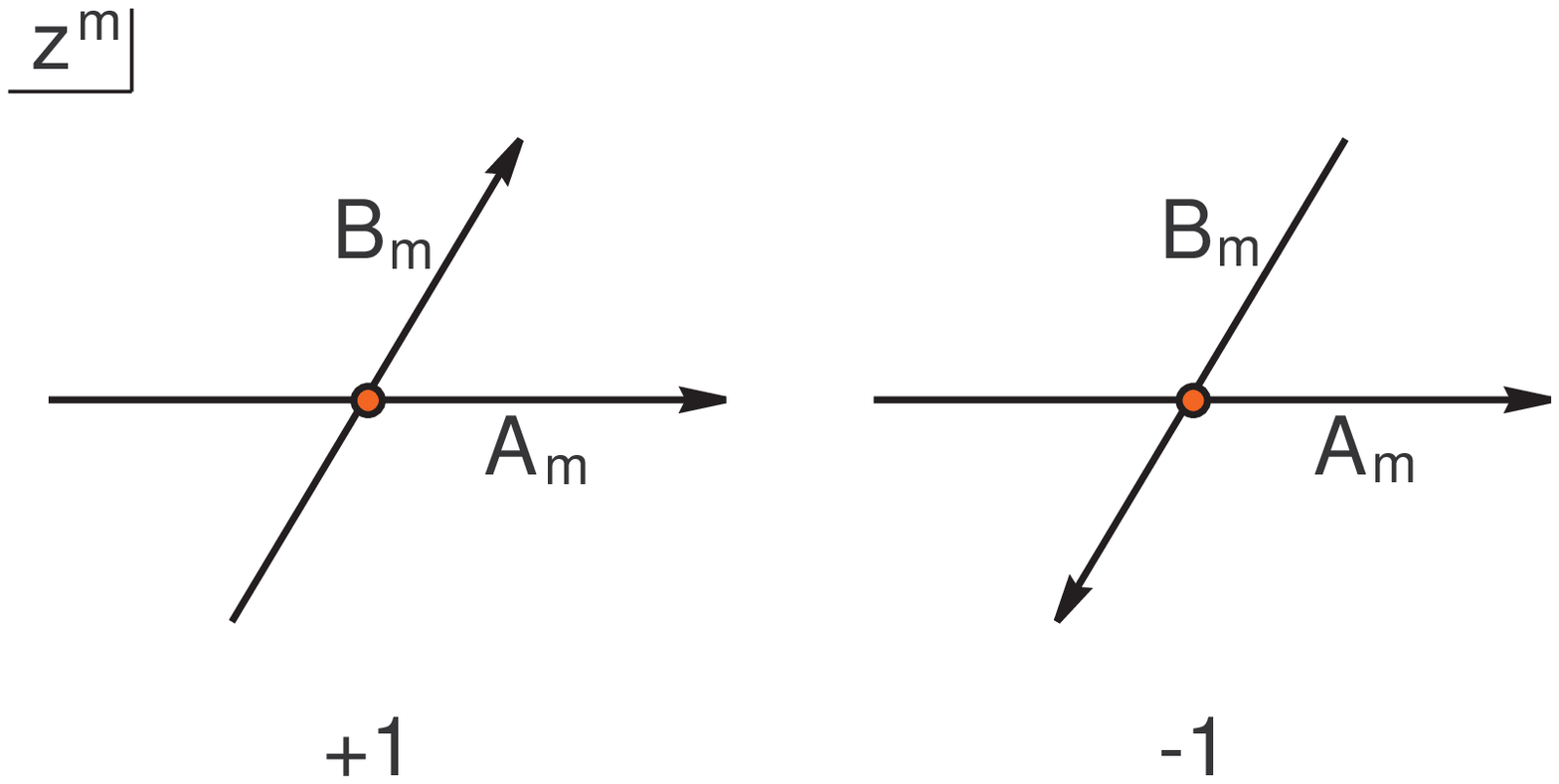,height=5cm}}
\caption{Contribution of a transversal intersection in the $z^m$
coordinate plane. \emph{Left:} $\int_{\IC} \hat{A}_m \wedge
\hat{B}_m = +1$. \emph{Right:} $\int_{\IC} \hat{A}_m \wedge
\hat{B}_m = -1$. }\label{intersect}}

We want to compute the intersection product of two $d$ real
dimensional submanifolds $A$ and $B$ in a $d$ complex dimensional
manifold, with all intersections between $A$ and $B$ transversal,
and $A$ and $B$ intersecting each complex complex coordinate plane
in a curve, as shown in fig.\ \ref{intersect}. That is, near an
intersection point the manifolds can be represented locally as
$A_1 \times A_2 \times \cdots A_d$ resp.\ $B_1 \times B_2 \times
\ldots \times B_d$, with the $A_m$ and $B_m$ oriented curves as in
fig.\ \ref{intersect}. Denoting the local Poincar\'e dual to $A_m$
by $\hat{A}_m$ (that is, locally, $\hat{A}_m = \delta(n) dn$, with
$n$ any coordinate normal to $A_m$) and similarly for $B_m$, we
find for the contribution of such an intersection point to the
intersection product
\begin{eqnarray}
 \int_{\IC^d} \hat{A}_1 \wedge \cdots \wedge \hat{A}_d \,
 \wedge \, \hat{B}_1 \wedge \cdots \wedge \hat{B}_d
 &=& (-1)^{d(d-1)/2} \int_{\IC^d} \hat{A}_1 \wedge \hat{B}_1  \, \wedge \cdots
 \wedge \, \hat{A}_d \wedge \hat{B}_d  \nonumber \\
 &=& (-1)^{d(d-1)/2} \int_{\IC} \hat{A}_1 \wedge
 \hat{B}_1 \, \cdots  \, \int_{\IC} \hat{A}_d \wedge \hat{B}_d \, .
 \nonumber
\end{eqnarray}
Where the separate integrals $\int \hat{A}_m \wedge \hat{B}_m$
evaluate to $+1$ or $-1$ according to the orientation conventions
shown in fig.\ \ref{intersect}. Finally, the total intersection
product is obtained by summing the contributions of all
intersection points.

\newcommand{\dgga}[1]{\href{http://xxx.lanl.gov/abs/dg-ga/#1}{\tt
dg-ga/#1}}
\renewcommand\baselinestretch{1.08}\normalsize

\newcommand{\mathdg}[1]{\href{http://xxx.lanl.gov/abs/math.DG/#1}{\tt
math.DG/#1}}
\renewcommand\baselinestretch{1.08}\normalsize

\newcommand{\mathag}[1]{\href{http://xxx.lanl.gov/abs/math.AG/#1}{\tt
math.AG/#1}}
\renewcommand\baselinestretch{1.08}

\end{document}